\documentclass[10pt,journal,compsoc]{IEEEtran}

\usepackage{amsmath,amsfonts}
\usepackage{url}
\usepackage[switch]{lineno}
\usepackage{setspace}
\usepackage[normalem]{ulem}
\usepackage{hyperref}
\usepackage{xspace}
\usepackage{xcolor}
\usepackage{graphicx}
\usepackage{enumitem}
\usepackage{fancyvrb}
\usepackage{listings}
\usepackage{amsthm}
\usepackage{booktabs}
\usepackage{multirow}
\usepackage{textcomp}
\usepackage{tcolorbox}
\usepackage{colortbl}
\usepackage{makecell}
\usepackage[ruled,vlined,linesnumbered]{algorithm2e}
\usepackage{listings}
\usepackage{algpseudocode}
\usepackage{pifont}

\newcommand{\aflnet}{{\sc AFLnet}\xspace}
\newcommand{\aflnetBold}{{\sc\bfseries AFLnet}\xspace}
\newcommand{\aflnetBlack}{{\sc AFLnetBlack}\xspace}
\newcommand{\aflnetDark}{{\sc AFLnetDark}\xspace}
\newcommand{\aflnetCode}{{\sc AFLnetCode}\xspace}
\newcommand{\aflnetCodeBold}{{\sc\bfseries AFLnetCode}\xspace}
\newcommand{\aflnetQueue}{{\sc AFLnetQueue}\xspace}
\newcommand{\aflnetQueueBold}{{\sc\bfseries AFLnetQueue}\xspace}
\newcommand{\aflnetIPSM}{{\sc AFLnetIPSM}\xspace}
\newcommand{\aflnetIPSMBold}{{\sc\bfseries AFLnetIPSM}\xspace}
\newcommand{\aflnetLegion}{{\sc AFLnetLegion}\xspace}
\newcommand{\legion}{{\sc Legion}\xspace}
\newcommand{\afl}{{\sc AFL}\xspace}

\newcommand{\nyxnet}{{\sc Nyx-Net}\xspace}
\newcommand{\chatafl}{{\sc ChatAFL}\xspace}

\newcommand{\profuzzbench}{{\sc ProFuzzBench}\xspace}
\newcommand{\sgfuzz}{{\sc SGFuzz}\xspace}
\newcommand{\nsfuzz}{{\sc NSFuzz}\xspace}

\newcommand{\stateafl}{{\sc StateAFL}\xspace}
\newcommand{\greenfuzz}{{\sc Green-Fuzz}\xspace}
\newcommand{\snapfuzz}{{\sc SnapFuzz}\xspace}

\newcommand{\sbf}{\textsc{Boo\-Fuzz}\xspace}
\newcommand{\boo}{\sbf}
\newcommand{\envfuzz}{{\textsc{$\mathcal{E}$fuzz}}\xspace}
\newcommand{\chaosafl}{{\sc ChaosAFL}\xspace}

\definecolor{deepblue}{rgb}{0,0,0.5}
\definecolor{deepred}{rgb}{0.6,0,0}
\definecolor{deepgreen}{rgb}{0,0.5,0}
\definecolor{halfgray}{gray}{0.55}
\definecolor{ipythonframe}{RGB}{207, 207, 207}
\definecolor[named]{ACMBlue}{cmyk}{1,0.1,0,0.1}
\definecolor[named]{ACMYellow}{cmyk}{0,0.16,1,0}
\definecolor[named]{ACMOrange}{cmyk}{0,0.42,1,0.01}
\definecolor[named]{ACMRed}{cmyk}{0,0.90,0.86,0}
\definecolor[named]{ACMLightBlue}{cmyk}{0.49,0.01,0,0}
\definecolor[named]{ACMGreen}{cmyk}{0.20,0,1,0.19}
\definecolor[named]{ACMPurple}{cmyk}{0.55,1,0,0.15}
\definecolor[named]{ACMDarkBlue}{cmyk}{1,0.58,0,0.21}

\hypersetup{colorlinks,
  linkcolor=ACMDarkBlue,
  citecolor=ACMPurple,
  urlcolor=ACMDarkBlue,
  filecolor=ACMDarkBlue}

\lstdefinestyle{base}{
  language=C,
  emptylines=1,
  breaklines=true,
  numbersep=15pt,
  basicstyle=\footnotesize\ttfamily\color{black},
  moredelim=**[is][\color{red}]{@}{@},
  linewidth=0.48\textwidth,
  xleftmargin=2.5em,
  framexleftmargin=1em,
}

\definecolor{mycolor}{rgb}{0.122, 0.435, 0.698}
\definecolor{gray1}{gray}{0.3}

\definecolor{darkgreen}{rgb}{0.0, 0.5, 0.0}
\definecolor{darkred}{rgb}{0.82, 0.1, 0.26}
\newcommand{\xmark}{\textcolor{darkred}{\ding{55}}}%

\definecolor{codegreen}{rgb}{0,0.6,0}
\definecolor{codegray}{rgb}{0.5,0.5,0.5}
\definecolor{codepurple}{rgb}{0.58,0,0.82}
\definecolor{backcolour}{rgb}{0.95,0.95,0.92}

\newcommand{\result}[1]{%
\begin{tcolorbox}[colframe=mycolor,boxrule=0.5pt,arc=4pt,
      left=6pt,right=6pt,top=6pt,bottom=6pt,boxsep=0pt,width=\columnwidth]%
      {#1}
\end{tcolorbox}%
}

\newcommand\hl{\bgroup\markoverwith
  {\textcolor{gray!30}{\rule[-.5ex]{2pt}{2.5ex}}}\ULon}
\newcommand{\MyComment}[1]{\Comment{\hl{#1}}}
\newcommand{\changes}[1]{\textcolor{black}{#1}}

\newcommand{\inlineSubSubSection}[1]{%
  \vspace{0.5\baselineskip}\noindent\sffamily\slshape{#1}\hspace{0.5em plus 0.5em minus 0.5em}\normalfont}

\begin{document}

\title{AFLNet Five Years Later: On Coverage-Guided Protocol Fuzzing}

\author{
    Ruijie Meng,
    Van-Thuan Pham,
    Marcel B{\"o}hme,
    and Abhik Roychoudhury%
    \IEEEcompsocitemizethanks{
        \IEEEcompsocthanksitem
        R. Meng is with the National University of Singapore, Singapore.
        E-mail: ruijie\_meng@u.nus.edu.
        \IEEEcompsocthanksitem
        V.T. Pham is with the University of Melbourne, Australia.
        E-mail: thuan.pham@unimelb.edu.au.
        \IEEEcompsocthanksitem
        M. B{\"o}hme is with the Max Planck Institute for Security and Privacy, Germany.
        Email-: marcel.boehme@acm.org.
        \IEEEcompsocthanksitem
        A. Roychoudhury is with the National University of Singapore, Singapore.
        E-mail: abhik@comp.nus.edu.sg.
    }
}


\IEEEtitleabstractindextext{%
\begin{abstract}
Protocol implementations are stateful which makes them difficult to test: Sending the same test input message twice might yield a different response every time. Our proposal to consider a sequence of messages as a seed for coverage-directed greybox fuzzing, to associate each message with the corresponding protocol state, and to maximize the coverage of both the state space and the code was first published in 2020 in a short tool demonstration paper. AFLNet was the first code- and state-coverage-guided protocol fuzzer; it used the response code as an indicator of the current protocol state. Over the past five years, the tool paper has gathered hundreds of citations, the code repository was forked almost 200 times and has seen over thirty pull requests from practitioners and researchers, and our initial proposal has been improved upon in many significant ways. In this paper, we first provide an extended discussion and a full empirical evaluation of the technical contributions of AFLNet and then reflect on the impact that our approach and our tool had in the past five years, on both the research and the practice of protocol fuzzing.
\end{abstract}

}

\maketitle

\IEEEraisesectionheading{\section{Introduction}}

It is critical to find security flaws in protocol implementations. Protocols are used by internet-facing servers to talk to each other or to clients in an effective and reliable manner. A \emph{protocol} specifies the exact sequence and structure of messages that can be exchanged between two or more online parties. However, this ability to talk to a server from anywhere in the world provides ample opportunities for remote code execution attacks. An attacker does not even require physical access to the machine. For instance, the famous Heartbleed vulnerability is a security flaw in OpenSSL, an implementation of the SSL/TLS protocol which promises secure communication.\footnote{See \url{http://heartbleed.com/}}

\begin{figure}[t] 
    \centering
    \includegraphics[width=0.9\columnwidth]{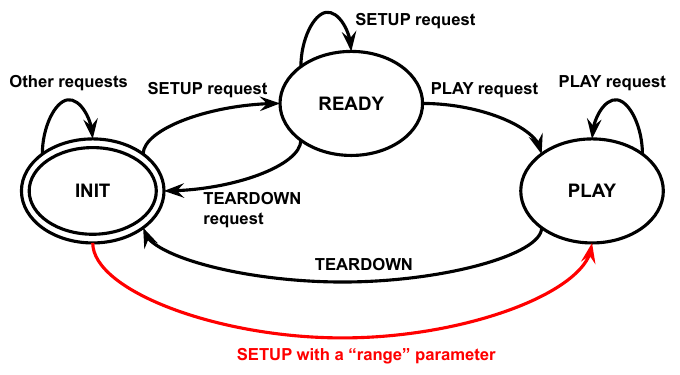}
    \caption{RTSP as implemented in Live555. There exists an unspecified shortcut between the INIT and PLAY state (shown in red).}
    \label{fig:live555}
\end{figure}

However, finding vulnerabilities in protocol implementations is also difficult. First, a server is stateful and message-driven. It takes a sequence of messages (a.k.a. requests) from a client, processes the messages, and sends appropriate responses. Yet, the \emph{implemented} protocol may not entirely correspond to the \emph{specified} protocol, making model-based fuzzing approaches \cite{peach, spike} less effective. For instance, as shown in Figure~\ref{fig:live555}, the Live555 streaming media server implements a state machine for the Real-Time Streaming Protocol (RTSP) that \changes{unintentionally introduces} an unspecified transition between the INIT and PLAY states (shown in red). Second, the server's response depends on both the current message and its internal state, which is influenced by earlier messages, posing challenges for vanilla coverage-guided greybox fuzzers like American Fuzzy Lop (\afl) and its extensions \cite{aflsmart, aflgo}. 

Before the extension to state-coverage, greybox fuzzers were primarily designed to test stateless programs (e.g., command line programs or libraries) where the same input would mostly produce the same output. If a generated input covered source code that was not previously covered, it was added to the set of input seeds for later fuzzing. If there was any program state, it would not be considered. Indeed, users of \afl were aware of these limitations and submitted several requests and questions for stateful fuzzing support to its developers' group~\cite{aflgroup}, as shown in \autoref{fig:quotes}.

\begin{figure}[t]
\result{
\emph{``One of the things that I struggle with is the limitation AFL seems to have, in that it only performs fuzzing with one input (a file). For many systems, such as network protocols, it would be useful if fuzzing could be done on a sequence of inputs. This sequence of inputs might be for example messages necessary to complete a handshake in TLS/TCP.''}\\[0.1cm]
\hspace*{\fill} - Paul (a member of the \afl's user group) \cite{aflgroup}
}
\vspace{-0.3cm}
\end{figure} 

\begin{figure}[t]
\result{
\emph{``I'm interested in doing something fairly non-traditional and definitely not currently supported by AFL. I would like to perform fuzzing of a large and complex external server that cannot easily be stripped down into small test cases.''}\\[0.1cm]
\hspace*{\fill} - Tim Newsham (a member of the \afl's user group) \cite{aflgroup}
}
\caption{Requests from AFL's users asking for stateful fuzzing support.}
\label{fig:quotes} 
\end{figure} 

In 2020, motivated by the aforementioned challenges of stateful network protocol fuzzing and the pressing need for an effective tool by researchers and practitioners, we introduced \aflnet \cite{aflnet}--the first code- and state-coverage-guided greybox fuzzer. However, an extended technical discussion and a full empirical evaluation of all its components in a full-length article have since been outstanding.

\aflnet integrates automated state model inference with coverage-guided fuzzing, allowing both to work in tandem: fuzzing generates new message sequences to cover new states and incrementally complete the state model. Meanwhile, the dynamically constructed state model helps drive fuzzing towards more critical code regions by leveraging both state coverage and code coverage information from the retained message sequences. With these advanced features, \aflnet successfully generated a random message sequence that discovered the hidden transition in the RTSP implementation of Live555 (see \autoref{fig:live555}), retained the sequence, and systematically evolved it to uncover a critical zero-day vulnerability (CVE-2019-7314, CVSS score: 9.8).
In March 2020, we released \aflnet as an open-source tool on GitHub: \textbf{\url{https://github.com/aflnet/aflnet}}.

Over the past five years, our tool has received tremendous attention from both the research community and industry. As of November 2024, it has garnered 872 stars on GitHub. It supports 17 protocols\footnote{List of protocols supported by \aflnet{}: RTSP, FTP, MQTT, DTLS, DNS, DICOM, SMTP, SSH, TLS, SIP, HTTP, IPP, TFTP, DHCP, SNTP, NTP, and SNMP.}, most of which were contributed by other researchers. 
\emph{Security researchers} have written experience reports and tutorials about the application of \aflnet to challenging targets such as as the 5G network \cite{nccgroup5g}, Internet of Things (IoT) \cite{matteraflnetbug1, matteraflnetbug2}, medical imaging applications \cite{dicomaflnet}, and automotive systems \cite{etaswebinar} highlighting its impact on practice. 
\emph{Researchers} have cited the short \aflnet tool demo paper hundreds of times (270+ according to Google Scholar), highlighting its impact on research. 
\emph{Educators} are introducing \aflnet as a coverage-guided protocol fuzzer to hundreds of Master's students at several universities, including the University of Melbourne and Carnegie Mellon University, highlighting its impact on education.

In this paper, we first provide an extended discussion and a full empirical evaluation of the technical contributions of \aflnet. 
We evaluate \aflnet in large-scale experiments on the widely-used ProFuzzBench benchmark \cite{profuzzbench} to provide researchers and practitioners with a deeper understanding of the capabilities of \aflnet and the effectiveness of each of its components. Specifically, we thoroughly analyze the effectiveness of state feedback both independently and in combination with traditional code coverage feedback. Additionally, we evaluate the impact of different seed-selection strategies implemented in \aflnet.
Based on these results, we offer practical guidance to \aflnet users on its optimal use cases and the most effective configurations to maximize their results. 

Finally, we reflect on the impact that our approach and our tool had in the past five years, on both the research and the practice of protocol fuzzing. This reflection not only illustrates \aflnet’s growing impact but also identifies open challenges and opportunities, shedding light on recent progress and promising new directions for future research in stateful network protocol fuzzing.

\section{Background and Motivation}

\subsection{Motivating Example: File Transfer Protocol (FTP)}

We begin with an informal introduction of the main concepts behind server communication and the terminology we are using in this paper. A \emph{server} is a software system that can be accessed remotely, e.g., via the Internet. A \emph{client} is a software system that uses the services which are provided by a server. In order for the client to use the services of a server, the \changes{client} must first establish a connection via a communication channel. A \emph{network socket} is an endpoint for sending or receiving data and can be identified by an IP address and a port.

In order to exchange information, both network participants send messages. A \emph{message} is a distinct data packet. A \emph{message sequence} is a vector of messages. A valid order of messages is governed by a protocol. The \emph{protocol} provides strict rules and regulations that determine how data is transmitted and ensures reliable communication between client and server. A message from the client is also called \emph{request} while a message from the server is called \emph{response}\footnote{\changes{In some protocols, e.g., mutual authentication in TLS, when the client is authenticating its identity to the server, the request comes from the server.}}. Each request may advance the server state, e.g., from initial state to authenticated. The \emph{server state} is a specific status of the server in the communication with the client.

\lstset{numbers=left, numberstyle=\tiny}
\begin{lstlisting}[frame=single, style=base, caption={Message exchange between an FTP client (red) and the LightFTP server (black) on the \texttt{control} channel.} \vspace{0.5cm}, label={lst:trace}]
220 LightFTP server v2.0a ready
@USER foo@
331 User foo OK. Password required
@PASS foo@
230 User logged in, proceed.
@MKD demo@
257 Directory created.
@CWD demo@
250 Requested file action okay, completed.
@STOR test.txt@
150 File status okay
226 Transfer complete
@LIST@
150 File status okay
226 Transfer complete
@QUIT@
221 Goodbye!
\end{lstlisting}

\begin{table*}[t]
    \setlength{\abovecaptionskip}{5pt}%
    \caption{Limitations of traditional fuzzing approaches in finding vulnerabilities in stateful protocols}
    \label{tab:approach_limitations}
    \centering
    \small
    \setlength\tabcolsep{3pt}
    \def\arraystretch{1.1}
     \begin{tabular}{
         >{\color{black}}m{0.13\linewidth}|
        >{\color{black}}m{0.27\linewidth}|
        >{\color{black}}m{0.17\linewidth}|
        >{\color{black}}m{0.38\linewidth}}
        \toprule
        {\textbf{Approach}} & \textbf{Description} & \textbf{Representative Tools} & 
        \textbf{Limitations} \\
        \hline
        \hline
        {\raggedright Coverage-guided greybox fuzzing (CGF)} & {\raggedright Leverage code coverage information to retain and prioritize interesting seeds generated by mutation operators} & {\raggedright AFL \cite{afl} and its extensions ~\cite{aflfast, aflgo, aflsmart}} & {\raggedright Neither know the server state information nor the required message structure or order to be sent} \\
        \hline
        {\raggedright Workaround solutions of coverage-guided greybox fuzzing} & {\raggedright (1) Write test harness for unit testing of specific program states of the server under test (SUT); \\ (2) Concatenate message structures into files and use them as seeds to do normal mutational file fuzzing} & {\raggedright For (1), libFuzzer \cite{libfuzzer}; \\ For (2), AFL \cite{afl}} & {\raggedright For (1), cannot thoroughly test the interactions / transitions between several program states; requires substantial manual effort to write a correct test harness; and cannot test the whole server whose source code is not available; \\ For (2), inefficiency and ineffectiveness in bug finding due to no knowledge of which message to focus on and no state transition information} \\
        \hline
        \raggedright Stateful blackbox fuzzing (SBF) & {\raggedright Traverse the given protocol model and leverage data models/grammars of messages to generate message sequences from scratch} & {\raggedright beSTORM~\cite{bestorm}, BooFuzz~\cite{boofuzz}, Peach~\cite{peach} and Sulley~\cite{sulley}} & {\raggedright Writing state models and data models involves much manual effort and expertise, which are also often error-prone; and learn nothing from past fuzzing execution} \\
        \bottomrule
    \end{tabular}
\end{table*} 

\autoref{lst:trace} shows an exchange of messages according to the File Transfer Protocol (FTP) between a client and LightFTP \cite{lightftp}, a server that implements FTP and is one of the subjects in our evaluation. The message sequence sent from the client is highlighted in red. FTP is the standard protocol for transferring files (RFC959~\cite{ftp}). FTP specifies that a client must first authenticate itself at the server. Only after successful authentication can the client issue other commands (i.e., transfer parameter commands and service commands). For each request message from the client, the FTP server replies with a response message containing a status code (e.g., 230 [login successful] or 430 [invalid user/pass]). The status code in the response ensures that client requests are acknowledged and informs the client about the current server state. 

Finding vulnerabilities in protocol implementations is challenging. First, a server is stateful and message-driven. It takes a sequence of messages (e.g., messages shown in \autoref{lst:trace} in red) from a client, handles the messages, and sends appropriate responses. In addition, a server features a massive state space that can be traversed effectively only with specific sequences of messages. For example, only after accepting the correct user name and password (i.e., \texttt{USER foo} and \texttt{PASS foo}), LightFTP can transition to the state where it can process the \texttt{MKD demo} command. Second, a server's response depends on both, the current message and the current internal server state which is controlled by a sequence of earlier messages. 

\subsection{Difficulties of Traditional Fuzzing Approaches}

To find vulnerabilities in stateful protocols, there are several challenges for state-of-the-art fuzzing approaches, like coverage-based greybox fuzzing (CGF)~\cite{afl,libfuzzer} and stateful blackbox fuzzing (SBF)~\cite{peach,boofuzz}. \changes{The details of each approach and their corresponding limitations are listed in \autoref{tab:approach_limitations}.} CGF is an effective automated vulnerability detection technique. It leverages code coverage information, obtained by lightweight code instrumentation, to retain and prioritize interesting seeds (e.g., input files) generated by mutation operators (e.g., bit flips and splicing) in an evolutionary fashion. However, vanilla CGF, like \afl and its extensions~\cite{aflfast, aflgo, aflsmart}, neither know the server state information nor the required structure or order of the messages to be sent. These CGF were mainly designed to test stateless programs (e.g., file processing programs) which always produce the same output for the same input. No state is maintained or taken into account. 
 
Developers only have workaround solutions to fuzz protocol implementations using current CGF approaches. They would need to write test harnesses for unit testing of specific program states of the server under test (SUT) \cite{libfuzzer} or to concatenate message sequences into files and use them as seeds to do normal mutational file fuzzing \cite{afl}. These two approaches have several drawbacks. While unit testing is effective at some specific program states, it may not be able to thoroughly test the interactions/transitions between several program states. Moreover, it usually requires a substantial effort to write a new test harness to maintain correct program states and avoid false positives. Importantly, it is not applicable for end-to-end fuzzing to test the whole server whose source code may not be available. 

Working on concatenated files leads to inefficiency and ineffectiveness in bug finding. First, for each fuzzing iteration, the whole selected seed file needs to be mutated. Given a file \texttt{$f$} which is constructed by concatenating a sequence of messages from \texttt{$m_1$} to \texttt{$m_n$}, CGF mutates the whole file \texttt{$f$} and treats all messages equally. Suppose a message \texttt{$m_i$} is the most interesting one (e.g., exploring it leads to higher code coverage and potential bugs), CGF repeats mutating uninteresting messages \texttt{$m_1$} to \texttt{$m_\texttt{i-1}$} before working on \texttt{$m_i$} and it has no knowledge to focus on \texttt{$m_i$}. Second, lacking state transition information, CGF could produce many invalid sequences of messages which are likely to be rejected by the SUT. 

Due to the aforementioned limitations of CGF on stateful server fuzzing, the most popular technique is still stateful blackbox fuzzing (SBF). Several SBF tools have been developed in both academia (e.g., Sulley, BooFuzz \cite{sulley, boofuzz}), and in the industry (e.g., Peach, beSTORM \cite{peach, bestorm}). These tools traverse a given protocol model, in the form of a finite state machine or a graph, and leverage data models/grammars of messages accepted at the states to generate (syntactically valid) message sequences and stress test the SUT. However, their effectiveness heavily depends on the completeness of the given state model and data model, which are normally written manually based on the developers' understanding of the protocol specification and the sample captured network traffic between the client and the server. These manually provided models may not capture correctly the protocols implemented inside the SUT. Protocol specifications contain hundreds of pages of prose-form text. Developers of implementations may misinterpret existing or add new states or transitions. Moreover, like other blackbox approaches, SBF does not retain interesting test cases for further fuzzing. More specifically, even though SBF could produce test cases leading to new interesting states, which have not been defined in its state model, SBF does not retain those for further explorations. It also does not update the state model at run-time.

To address the aforementioned limitations of current CGF and SBF approaches, we introduce \aflnet--the first stateful CGF (SCGF) tool. \aflnet is an evolutionary mutation-based fuzzer that leverages code as well as state feedback to efficiently and systematically explore the code and state space of a protocol implementation. In our setting, \aflnet acts as a client while the protocol server acts as the fuzz target. \aflnet makes automated state model inferencing and coverage-guided fuzzing work hand in hand; fuzzing helps to generate new message sequences to cover new states and make the state model gradually more complete. Meanwhile, the dynamically constructed state model helps to drive the fuzzing towards more important code parts by using both the state coverage and code coverage information of the retained message sequences. 

\section{The \aflnetBold Approach}\label{sec:sgf}

\aflnet is a network-enabled stateful greybox fuzzer that leverages additional state feedback from the server along with the code feedback to boost the coverage of a protocol implementation. \autoref{alg:sgf} provides a procedural overview. The \emph{input} is the server program under test $\mathcal{P}$, an initial (potentially empty) draft of the implemented protocol state machine (IPSM) $S$, and the actual, recorded network traffic $T$ between a client and $\mathcal{P}$. The traffic $T$ can be recorded traces from multiple sessions. The \emph{output} is a set of error-revealing message sequences $C_\text{\xmark}$ and the IPSM $S$ that has been augmented throughout the fuzzing campaign.

\begin{algorithm}[t]
    \caption{Stateful Network Protocol Fuzzing \label{alg:sgf}}
    \DontPrintSemicolon
    \small
    \newcommand\mycommfont[1]{\ttfamily\textcolor{blue}{#1}}
    \SetCommentSty{mycommfont}
    \SetNoFillComment
    \setstretch{1.1}
    \SetKwProg{Fn}{func}{:}{}
    \SetKwInOut{Input}{Input}
    \SetKwInOut{Output}{Output}
    \Input{Server program $\mathcal{P}$, Sniffer traces $T$, IPSM $S$}
    \Output{Crashes $C_\text{\xmark}$, Corpus $C$, and IPSM $S$}
    Corpus $C \leftarrow \emptyset$; \  Crashes $C_\text{\xmark} \leftarrow \emptyset$; Bitmap $B \leftarrow \emptyset$  \;
    \For(\MyComment{\small{Pre-processing Phase}}){each trace $t\in T$}{
        Sequence $M \leftarrow \text{{\emph{parse}}}(t)$ \;
        Corpus   $C \leftarrow C \cup \{M\}$ \;
        Response $R \leftarrow \text{\emph{send}}(\mathcal{P}, M, B)$ \;
        IPSM     $S \leftarrow \text{\emph{updateIPSM}}(S, R)$ \;
    }
    LastPathTime $lpt$ $\leftarrow$ $\mathit{cur\_time}$ \;
    \Repeat(\MyComment{\small{Fuzzing Phase}}){\textup{timeout reached or abort}}{
        \uIf{$ (cur\_time - lpt) > MaxTimeGap$}{
            State $s \leftarrow \text{\emph{choose\_state}}(S)$ \;
            Sequence $M \leftarrow \text{\emph{choose\_sequence\_to\_state}}(C, s)$ \;
            $\langle M_1, M_2, M_3\rangle \leftarrow M$ \parbox[t]{.58\linewidth}{(i.e., split $M$ in subsequences such that $M_1$ is the message sequence to drive $\mathcal{P}$ to arrive at state $s$, and message sequence $M_2$ is selected to be mutated)} 
        }
        \Else(\MyComment{\small{Interleaving Seed Selection}}){
            Sequence $M \leftarrow \text{\emph{choose\_sequence\_from\_queue}}(C)$ \;
            $\langle M_1, M_2, M_3\rangle \leftarrow M$ \parbox[t]{.58\linewidth}{(i.e., $\mathit{randomly}$ select subsequence $M_2$ to be mutated)}
        }
        \For{$i$ from 1 to $\text{\emph{energy}}(M)$}{
            Sequence $M' \leftarrow \langle M_1, \text{\emph{mutate}}(M_2), M_3\rangle$ \;
            Response $R$ $\leftarrow$ $\text{\emph{send}}(\mathcal{P}, M', B)$ \;
            \uIf{$\mathcal{P}$ has crashed}{
                Crashes $C_\text{\xmark} \leftarrow C_\text{\xmark} \cup \{M'\}$ \;
                LastPathTime $lpt$ $\leftarrow$ $\mathit{cur\_time}$ \;
            }
            \ElseIf{$\mathit{is\_interesting}$($M', B$)}{
                Corpus $C \leftarrow C \cup \{M'\}$ \;
                IPSM   $S \leftarrow \text{\emph{updateIPSM}}(S, R)$ \;
                LastPathTime $lpt$ $\leftarrow$ $\mathit{cur\_time}$ \;
            }
        }
    }
    
\end{algorithm}

\aflnet starts with a pre-processing phase (Lines~1--6). Given the server $\mathcal{P}$, the initial IPSM $S$, and the recorded network traffic $T$, \aflnet constructs the initial seed corpus $C$ and adds state transitions observed in $T$ to $S$. In order to construct $C$, each trace $t\in T$ of recorded network traffic is parsed into the corresponding message sequence $M$, which is then added to $C$ and sent to the server $\mathcal{P}$. The details of the recording and replay of message sequences (incl. the \emph{parse} and \emph{send} methods) are discussed in \autoref{sec:recordreplay}. From each server response $R$, the exercised state transitions are extracted. States and transitions that have been observed are added to the IPSM $S$ if they do not already exist. If no IPSM is given as input, $S$ is initialized as a directed graph without nodes and edges. Our lightweight protocol learning (incl. the \emph{updateIPSM} method) is elaborated in \autoref{sec:protlearn}.
 
In each iteration (Lines~8--26), \aflnet generates several new sequences based on the selected seed sequence. During seed selection, \aflnet interleaves AFL's original strategy, which relies on the order of the seed queue (Lines~14--15), with a strategy based on the state heuristics in the IPSM $S$ (Lines~10--12). Specifically, \aflnet starts lightweight seed selection based on the seed queue, and switches to the heavy state-heuristic-based strategy when the fuzzer cannot find ``interesting'' sequences within the allowed time gap. 
The former strategy is the same as that used in \afl. In the latter seed-selection strategy, \aflnet selects a progressive state $s\in S$ and a sequence $M\in C$ exercising $s$ to steer the fuzzer towards more progressive regions in the server state space (\autoref{sec:steer}; incl. \emph{choose\_state} and \emph{choose\_sequence\_to\_state}). 

The selected sequence is assigned \changes{an amount of} energy based on the default power schedule of the greybox fuzzer~\cite{aflfast} and systematically mutated (\autoref{sec:mutate}; incl. \emph{mutate} and \emph{is\_interesting}). Crash-triggering sequences are added to the crashing corpus $C_\text{\xmark}$, ``interesting'' sequences are added to the normal corpus $C$, and all other generated sequences are discarded (Line~19--25). A sequence \textit{M} is considered as \emph{interesting} if the new state(s) or state transition(s) have been observed in the server response \textit{R} for \textit{M}, or \changes{if} \textit{M} covers new branches in the server's source code.

\subsection{Recording and Replay for Fuzzing}\label{sec:recordreplay}
In order to facilitate mutational fuzzing for message sequences, we first need to develop the capability to record and replay message sequences. In order to \emph{record} a realistic message exchange between \changes{the} client and \changes{the} server, a network sniffer can be used. A \emph{network sniffer} captures network traffic for a specified period of time. For instance, we can use \texttt{tcpdump}\footnote{\url{https://www.tcpdump.org/pcap.html}} to capture the traffic from a user-generated FTP session. The sniffer records the entire network traffic that can be filtered automatically. The relevant message exchange can be extracted using a packet analyzer. A \emph{packet analyzer} can identify and distinguish different message exchanges between different nodes in the network. For instance, we used the packet analyzer \texttt{Wireshark}\footnote{\url{https://www.wireshark.org/}} to automatically extract the sequence of FTP requests.

\begin{figure}[ht]\centering
\includegraphics[width=1\columnwidth]{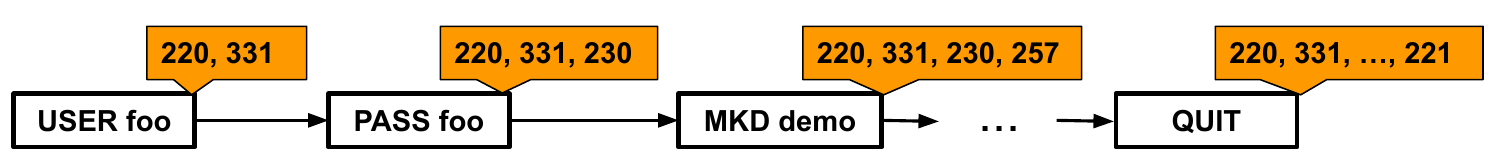}
\caption{An annotated FTP message sequence processed for mutational fuzzing (from the sniffer trace in \autoref{lst:trace}).}
\label{fig:msgSeq} 
\end{figure}

To generate the initial corpus of message sequences $C$, \aflnet parses the filtered sniffer traces $T$ (i.e., \emph{parse} in \autoref{alg:sgf}). The objective of the parser is to identify the start and end of a message in the filtered trace. This can be done with a packet analyzer such as \texttt{Wireshark}. In our case, we implemented a lightweight method that finds the header and terminator of a message as specified in the given protocol. For instance, each FTP message starts with a valid FTP command (e.g., \texttt{USER}, \texttt{PASS}) and is terminated with a carriage return followed by a line feed character (i.e., \texttt{0x0D0A}). Moreover, \aflnet associates with each message in the sequence the corresponding server state transitions (cf. \autoref{fig:msgSeq}). This is done by sending the messages and parsing the responses one by one. 

To \emph{replay} a message sequence (i.e., \emph{send} in \autoref{alg:sgf}), \aflnet acts as a client. In our setting, the server provides network sockets and the fuzzer can connect to those. After the server is started and the connection is established, \aflnet can proceed to replay message sequences. For each request $m$ in the sequence, (i) the request $m$ is sent to the server, (ii) a delay is introduced, and (iii) the response is received. 
The delay is required because many servers (incl. the LightFTP server) stop the message exchange if a new request arrives before the server's response has been received. When the entire sequence is executed, the connection is closed, and the server is terminated. We suggest terminating the server (or at least the ongoing session) because this will reset any accumulated state. The next message sequence can start from the same initial state.

\vspace{0.2cm}
\subsection{Lightweight Protocol Learning}\label{sec:protlearn}
\changes{We refer to the} directed labelled graph which reconciles all state transitions that have been observed throughout the fuzzing campaign \changes{as the} \emph{implemented protocol state machine} (IPSM). Each node represents a \emph{state}. Each directed edge represents a \emph{state transition}. The edge is annotated with a request and response message. If there is an edge between two states $s_1$ and $s_2$, and the server is currently in state $s_1$ and receives a request that matches the one in the edge label, then the server sends a response that matches the one in the edge label and transitions to state $s_2$. An example illustrating the utility of the IPSM is shown in \autoref{fig:live555}.

The IPSM represents the current and potentially incomplete view of the protocol state machine that has actually been implemented. The purpose of the fuzzer is to generate message sequences that discover \emph{new} state transitions. This in turn iteratively increases the completeness of the IPSM w.r.t. actual state machine.

After sending a request sequence $M\in C$, the network-enabled fuzzer receives a response sequence $R$.  From $R$, \aflnet extracts the sequence of state transitions. 
We assume \emph{determinism}, i.e., executing $M$ several times always produces the same sequence of state transitions.
Each state should be uniquely identifiable. For many protocols, the response contains information about the current server state. For instance, we can use the FTP status code in the server response to quickly identify the server state (e.g., 230 [login successful]). If no (detailed) state information is normally available in $R$, we suggest to instrument the program such that the server function that handles a certain state also prints the associated state ID. Such instrumentation is sensible when fuzzing in-house or open-source protocol implementations.

In order to augment the IPSM $S$ (cf. \emph{updateIPSM} in \autoref{alg:sgf}), the nodes and edges are added for states and state transitions that have not been observed previously (i.e., before sending $m\in M$). 
For each existing or new state $s\in S$, \aflnet records the number of times a mutated message sequence has executed $s$ (\texttt{\#fuzz}), the number of times $s$ has been selected for fuzzing (\texttt{\#selected}), and the number of coverage-increasing message sequences that have been added to $C$ after selecting $s$ for fuzzing (\texttt{\#paths}). This statistical information is used for steering the fuzzer towards more progressive regions of the state space. In turn, the boosted fuzzer should enable a more efficient augmentation of the IPSM.

\vspace{0.2cm}
\subsection{Steering the Fuzzer to Progressive States}\label{sec:steer}
In order to steer the fuzzer towards more progressive regions in the state space, \aflnet chooses message sequences $M\in C$ to mutate that exercise one of the more progressive states and that is more likely to increase coverage (Lines~10--12 in \autoref{alg:sgf}).

\textit{Choosing a state.}
In each iteration, \aflnet selects a server state $s\in S$ in the IPSM $S$ to focus on (cf. \emph{choose\_state} in \autoref{alg:sgf}). \aflnet uses several heuristics that can be computed from the statistical data available in the learned IPSM. To identify \emph{fuzzer blind spots}, i.e., rarely exercised states, \aflnet chooses a state $s$ with a probability that is inversely proportional to the proportion of mutated message sequences that have executed $s$ (\texttt{\#fuzz}). \aflnet chooses \emph{recently discovered states} $s$ with higher priority by prioritizing states that have been rarely chosen for fuzzing (\texttt{\#selected}). In order to maximize the probability of discovering new state transitions, \aflnet chooses a state $s$ with higher priority that has been particularly successful in contributing to an increased code or state coverage when they were previously selected (\texttt{\#paths}).

\textit{Choosing a message sequence.} In each iteration, given the selected state $s\in S$, \aflnet selects a message sequence $M\in C$ from the corpus $C$ that exercises $s$ (cf. \emph{choose\_sequence\_to\_state} in \autoref{alg:sgf}). We leverage the original selection strategy that is provided by the greybox fuzzer but on the \emph{reduced} corpus of sequences that exercise the selected state $s$. For instance, classically AFL prioritizes shorter seeds in the corpus that execute quicker. Our modified strategy first filters only sequences that execute $s$. Shorter and quicker sequences that have reached more states are prioritized. 

\textit{Assigning energy.} In greybox fuzzing, the \emph{energy} of a seed input determines how many new inputs are generated from the given seed input the next time it is chosen (cf. \emph{energy} in \autoref{alg:sgf}). For instance, the AFL coverage-based greybox fuzzer \cite{afl} assigns more energy to a seed that executes faster and that is shorter. \aflnet leverages the default power schedule of the greybox fuzzer. A \emph{power schedule} is the mechanism that assigns the energy of a seed.

\subsection{Mutating a Message Sequence}\label{sec:mutate}

\aflnet is \emph{mutation-based fuzzer}, i.e., a seed message sequence is chosen from a corpus and mutated to generate new sequences. There are several advantages over existing \emph{generation-based approaches} which generate new message sequences from scratch. 
First, a mutation-based approach can leverage a valid trace of real network traffic to generate new sequences that are likely valid---albeit entirely without a protocol specification. In contrast, a generation-based approach \cite{sulley,spike,boofuzz} requires a detailed protocol specification, including concrete message templates and the protocol state machine. Hence, \boo \cite{boofuzz}, a generation-based approach, does not discover the unspecified state transition in \autoref{fig:live555} in page~\pageref{fig:live555}. 
Second, a mutation-based approach \changes{allows the fuzzer to evolve} a corpus of particularly interesting message sequences. Generated sequences that have led to the discovery of new states, state transitions, or program branches are added to the corpus for further fuzzing. This evolutionary approach is the secret sauce of the tremendous recent success of coverage-based greybox fuzzing.

Given a state $s$ and a message sequence $M$, \aflnet generates a new sequence $M'$ by mutation (cf. Line 17 in \autoref{alg:sgf}). In order to ensure that the mutated sequence $M'$ still exercises the chosen state $s$, \aflnet splits the original sequence $M$ into three parts: 
\begin{itemize}[itemsep=5pt]
  \item The \emph{prefix} $M_1$ is required to reach the selected state $s$. The prefix is identified using the state annotations (cf. \autoref{fig:msgSeq}). If $M=\langle m_1, \ldots, m_n\rangle$, then $M_1=\langle m_1,\ldots,m_i\rangle$ such that $s$ is  observed for the first time when message $m_i$ is sent, i.e.
  $s\not\in\text{\emph{states}}(\text{\emph{send}}(\mathcal{P},\langle m_1, .., m_{i-1}\rangle))\wedge s\in\text{\emph{states}}(\text{\emph{send}}(\mathcal{P},M_1))$.   

  \item The \emph{candidate subsequence} $M_2$ contains all messages that can be executed after $M_1$ while still \emph{remaining} in state $s$. In other words, $M_2$ is the longest subsequence $\langle M_1,M_2\rangle \subseteq M$, such that $\text{\emph{states}}(\text{\emph{send}}(\mathcal{P}, M_1)) = \text{\emph{states}}(\text{\emph{send}}(\mathcal{P}, \langle M_1, M_2\rangle))$.
  
  \item The \emph{suffix} $M_3$ is simply the left-over subsequence such that $\langle M_1, M_2, M_3\rangle = M$.
\end{itemize}

The mutated message sequence $M'=\langle M_1,mutate(M_2), M_3\rangle$. By maintaining the original subsequence $M_1$, $M'$ will still reach the state $s$ which is the state that the fuzzer is currently focusing on. The mutated candidate subsequence $mutate(M_2)$ produces an alternative sequence of messages \emph{upon} the progressive state $s$. In our initial experiments, we observed that the alternative requests may not be observable ``now'', but propagate to later responses. Hence, \aflnet continues with the execution of the original suffix $M_3$.

\aflnet offers several \emph{protocol-aware mutation operators} to modify the candidate subsequence (cf. \emph{mutate} in \autoref{alg:sgf}). From the corpus $C$ of message sequences, \aflnet produces a pool of messages. The \emph{message pool} is a collection of actual messages from network sniffer traces (plus generated messages) that can be added or substituted into existing message sequences $M\in C$. In order to mutate the candidate sequence $M_2$, \aflnet supports the replacement, insertion, duplication, and deletion of messages. In addition to these protocol-aware mutation operators, \aflnet uses the common byte-level operators that are known from greybox fuzzing, such as bit flipping, and the substitution, insertion, or deletion of blocks of bytes. The mutations are \emph{stacked}, i.e., several protocol-aware and byte-level mutation operators are applied to generate the mutated candidate sequence. The mutations affect the start and end indices of the mutated and any subsequent message in the sequence. Hence, the index annotations are updated accordingly. 

\subsection{Selecting Interesting Message Sequences}

After applying protocol-aware mutations on the selected message sequence $M$, \aflnet generates a new message sequence $M'$ and sends it to the server under test to investigate whether $M'$ is ``interesting'' (Line 18 in \autoref{alg:sgf}). A sequence is considered as \emph{interesting} if the server response contains new states or state transitions that have not previously been observed (i.e., they are not recorded in the IPSM $S$); a sequence is interesting also if it covers new branches in the server's source code.

To select ``interesting'' message sequences, \aflnet records both state coverage and branch coverage in the same bitmap $B$ during program execution. In contrast, \afl only records the branch coverage in the bitmap. While hitting a branch $B_1 \rightarrow B_2$, \afl computes the map index using \autoref{eq:code_cov} (where \emph{prev\_loc} and \emph{cur\_loc} are branch keys of the basic block of $B_1$ and $B_2$) and increments the corresponding value for this index by 1. However, based on our observation, this bitmap often has many empty entries in the default bitmap size (MAP\_SIZE), which provides the chance to maintain additional state coverage within the same bitmap as well. 

\begin{equation} \label{eq:code_cov}
\begin{gathered}
    map\_index = cur\_loc \oplus (prev\_loc \gg 1) 
\end{gathered}
\end{equation}

\begin{figure*}[ t]\centering
\includegraphics[width=\textwidth]{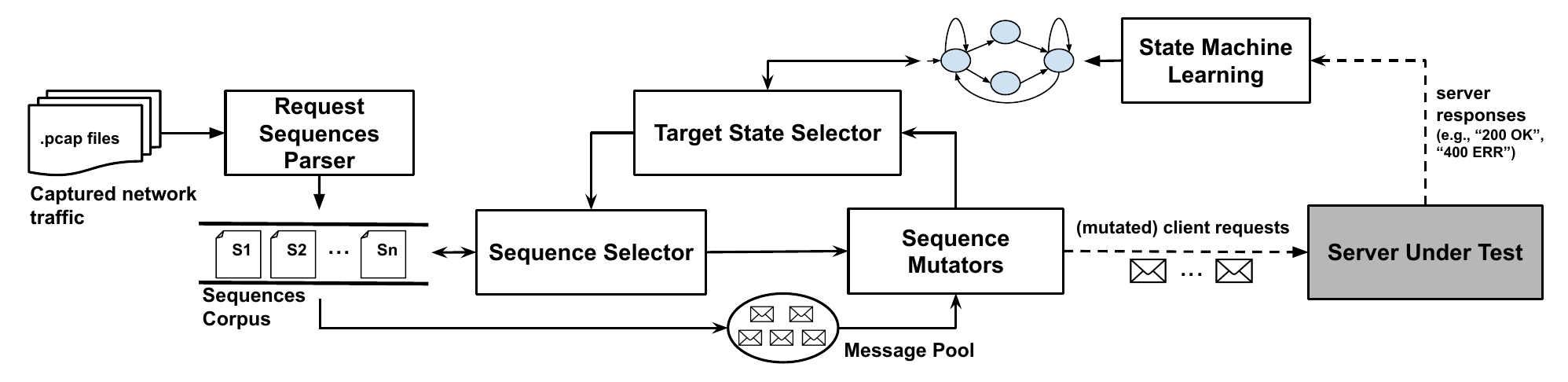}
\vspace{-0.6cm}
\caption{Architecture and Implementation of Stateful Greybox Fuzzing \changes{in} \aflnet.}
\label{fig:aflnet}
\end{figure*}

To achieve this, \aflnet shifts the code coverage to the left of the bitmap by a certain number of elements SHIFT\_SIZE, \changes{where SHIFT\_SIZE is the number of the bitmap entries reserved to store state coverage information). The map index for code branches is then computed} using \autoref{eq:new_code_cov}.

\begin{equation} \label{eq:new_code_cov}
\begin{gathered}
    map\_index = (cur\_loc \oplus (prev\_loc \gg 1)) ~\%~ {\rm (MAP\_SIZE -} \\ {\rm SHIFT\_SIZE) + SHIFT\_SIZE}
\end{gathered}
\end{equation}

Following this adjustment, \aflnet reserves the bitmap space with SHIFT\_SIZE many elements to state coverage. While observing a state transition $S_1 \rightarrow S_2$, \aflnet computes the map index for this state transition using \autoref{eq:state_cov}, where \emph{prev\_state} and \emph{cur\_state} are state keys by numbering raw states $S_1$ and $S_2$ starting from the number 1, \changes{and STATE\_SIZE is the maximum number of states expected to be observed at the end of fuzzing campaign. The value of the corresponding index is then incremented by 1.}  

\begin{equation} \label{eq:state_cov}
\begin{gathered}
    map\_index = (prev\_state \times {\rm STATE\_SIZE} + cur\_state) \\ ~\% ~ {\rm SHIFT\_SIZE}
\end{gathered}
\end{equation}

Based on the maintained bitmap, \aflnet selects ``interesting'' message sequences that hit new bitmap entries or increase the hit count (cf. \emph{is\_interesting} in \autoref{alg:sgf}). The interesting message sequences are saved into the seed corpus $C$ for further examination (Line 23). Meanwhile (Lines~24--25), \aflnet updates the states in IPSM $S$, and the time to find interesting paths, which facilitates tracking whether \aflnet enters the coverage plateau. 

\section{Implementation}\label{sec:impl}

We implemented our prototype \aflnet as an extension of the popular and successful greybox fuzzer \afl~\cite{afl,aflbugs}.
The architecture of \aflnet is shown in \autoref{fig:aflnet}. To facilitate communication with the server, we first enabled network communication over sockets, which is not supported by the vanilla \afl. \aflnet supports two channels, one to send and one to receive messages from the \textbf{Server Under Test}.
\aflnet uses standard C Socket APIs (i.e., $connect, poll, send$, and $recv$)\footnote{\url{http://man7.org/linux/man-pages/man2/socket.2.html}} to implement this feature. To ensure proper synchronization between \aflnet and the server under test, we added delays between requests (see \autoref{sec:recordreplay}). Otherwise, several server implementations drop the connection if a new message is received before the response is sent and acknowledged. To minimize the delay, we used the Linux \texttt{poll} API to monitor the status of both outgoing and incoming buffers. In our experiments, this led to a substantial speed up (3$\times$) compared to a static delay.

The \textbf{Request Sequences Parser} takes the \texttt{pcap} files containing the captured network traffic and produces the initial corpus of message sequences (cf. \autoref{sec:recordreplay}). \aflnet uses protocol-specific information of the message structure to extract individual requests, in correct order, from the captured network traffic. In order to reduce the onus on the developer, \aflnet only requires to specify a mechanism to identify message boundaries (i.e., start and end of individual request messages). For instance, for the four protocols in our evaluation, we implemented the method to extract request sequences (and to parse the state information from the server responses) into only 200 lines of C code. 

The \textbf{State Machine Learner} takes the server responses and augments the implemented protocol state machine (IPSM) with newly observed states and transitions. \aflnet reads the server response into a byte buffer, extracts the status code as specified in the protocol, and determines the executed state (transitions). To represent the IPSM, \aflnet uses the Graphviz graph libary\footnote{\url{https://www.graphviz.org/pdf/libguide.pdf}} and the Collections-C\footnote{\url{https://github.com/srdja/Collections-C}}, which supports high-level data structures like HashMap and List. These libraries are used to construct the state machine, and associate state-specific information. The GraphViz library can render the entire state machine as an image file. This allows users of \aflnet to understand intuitively the fuzzer progress in terms of the coverage of the state space.

The \textbf{Target State Selector} takes information from the IPSM to select \changes{the state that} \aflnet should focus on next. \afl implements the \emph{seed corpus} (here, containing message sequences) as a linked list of queue entries. A \emph{queue entry} is the data structure containing pertinent information about the seed input. In addition, \aflnet maintains a \emph{state corpus} which consists of (i) a list of \emph{state entries}, i.e., a data structure containing pertinent state information, and (ii)~a hashmap which maps a state identifier to a list of queue entries exercising the state corresponding to the state identifier. \changes{Both the} target state selector and the \textbf{Sequence Selector} leverage the state corpus. The \textbf{Sequence Mutator} augments AFL's \texttt{fuzz\_one} method with protocol-aware mutation operators. 

\begin{figure}[ht]
\centering
\includegraphics[width=0.9\columnwidth]{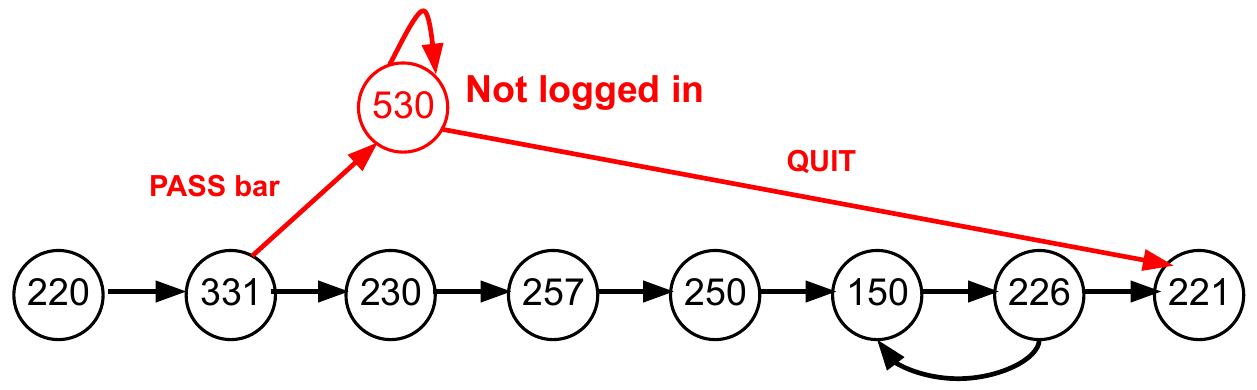}
\caption{Learning example of the implemented protocol state machine (IPSM) from the LightFTP Server. }
\label{fig:ipsm_example} 
\end{figure}

\begin{figure}[ht]\centering
\includegraphics[width=\columnwidth]{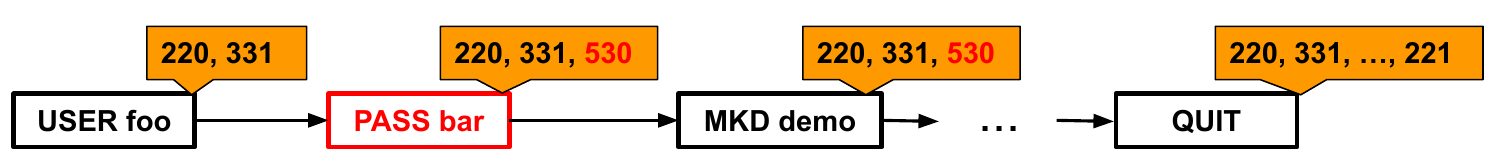}
\caption{A sample mutated sequence if the state 331 (User OK) and the message sequence in \autoref{fig:msgSeq} have been chosen.}
\label{fig:seq_example} 
\end{figure}

Now we illustrate how all these components of \aflnet work together to fuzz the LightFTP server. Suppose \aflnet starts with only one pcap file containing the network traffic as shown in \autoref{lst:trace}. First, \textbf{Request Sequence Parser} parses the pcap file to generate a single sequence (as visualized in \autoref{fig:msgSeq}) and \changes{saves} it into the corpus $C$. At the same time, \textbf{State Machine Learning} constructs the initial IPSM based on the response codes; this initial IPSM contains black nodes and transitions in \autoref{fig:ipsm_example}. Suppose that \textbf{Target State Selector} selects state \texttt{331} (``\texttt{USER foo OK}'') as the target state\changes{.} \textbf{Sequence Selector} will then randomly select a sequence from the sequence corpus $C$, which contains only one sequence at this moment. Afterward, \textbf{Sequence Mutators} identifies the sequence prefix (``USER foo'' request), the candidate subsequence (``\texttt{PASS foo}'' request), and the remaining subsequence as the suffix. By mutating the candidate subsequence using stacked mutators, \textbf{Sequence Mutators} may generate a wrong password request (``\texttt{PASS bar}'') leading to an error state (\texttt{530} \texttt{Not logged in}). Following this wrong password, it replays the suffix (e.g., ``\texttt{MKD demo}'', ``\texttt{CWD demo}'') leading to a loop in the state \texttt{530} because all these commands are not allowed before successful authentication. Finally, the ``\texttt{QUIT}" request is sent, and the server exits. Since the generated test sequence (as visualized in \autoref{fig:seq_example}) covers new state and state transitions (as highlighted in red in \autoref{fig:ipsm_example}), it is added into the corpus $C$ and the IPSM. 
\section{Evaluation} \label{sec:eva}

This evaluation seeks to analyze the contribution of each algorithmic component embedded within \aflnet.
To this end, we design three research questions to be covered in the following:

\begin{description}
\item [\textbf{RQ.1}] \textbf{How effective is state feedback alone in guiding the fuzzing campaign?} This research question aims to evaluate whether state feedback alone is effective in guiding the fuzzing campaign when code feedback is unavailable. We seek to measure this performance and examine the potential for extending \aflnet to scenarios such as fuzzing remote servers. 

\begin{figure*}[ht]
\setlength{\abovecaptionskip}{5pt}
  \centering
  \includegraphics[page=1, trim= 1.05in 1.2in 1.05in 0.6in, clip, scale=0.65]{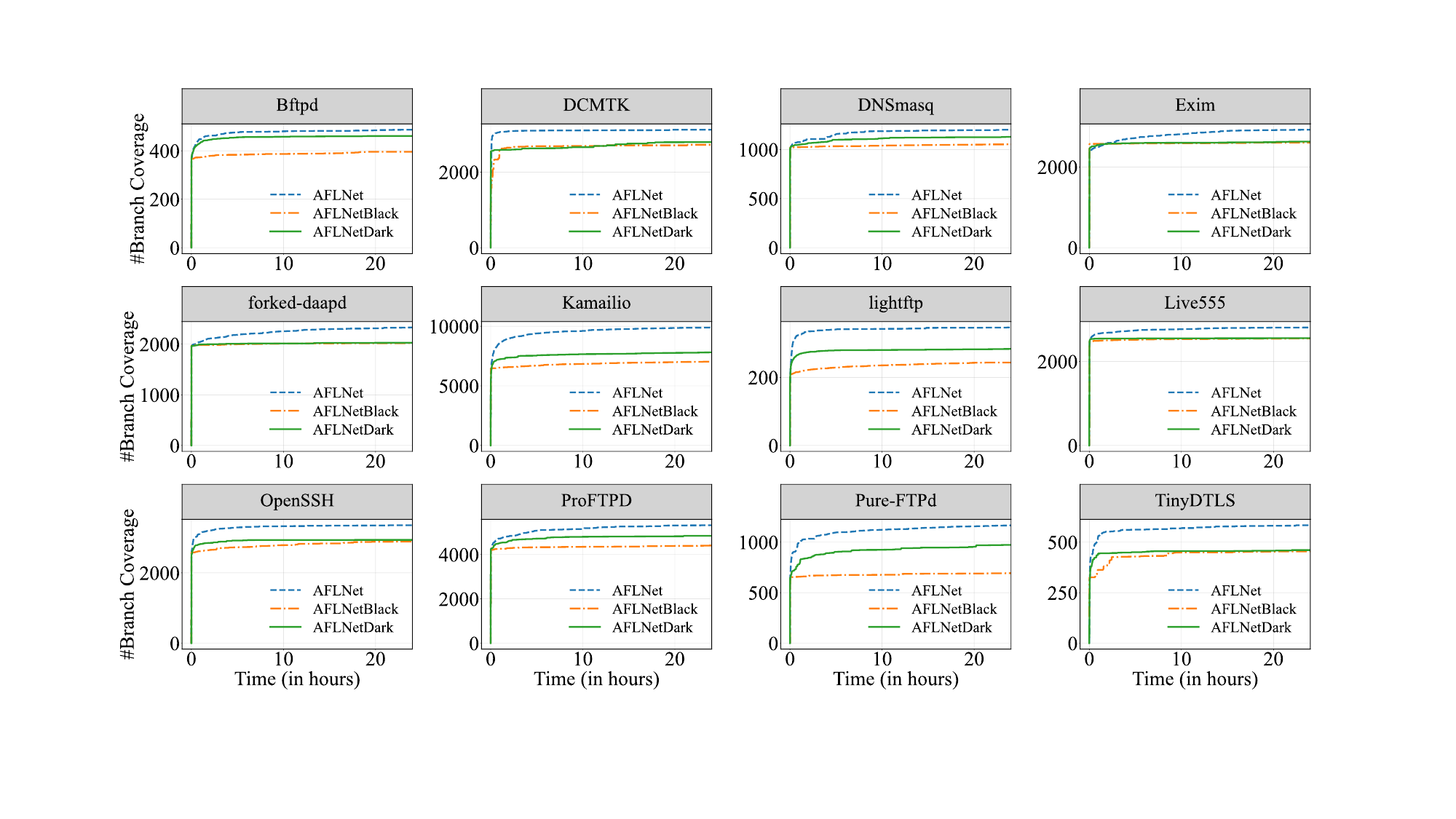}
  \caption{ Code branch covered over time by \aflnet, \aflnetDark and \aflnetBlack across 10 runs of 24 hours on the \profuzzbench subjects.}
  \label{fig:coverage-trend}
\end{figure*}

\item [\textbf{RQ.2}] \textbf{Can state feedback enhance the fuzzing effectiveness alongside code feedback?} This research question aims to evaluate the extent to which fuzzing effectiveness can be improved by incorporating additional state feedback. We expect the fuzzer to cover more code with the inclusion of state feedback.

\item [\textbf{RQ.3}] \textbf{What is the impact of different seed-selection strategies?} In the default configuration, \aflnet adopts an interleaving seed-selection strategy that alternates between the order in the seed queue and the state heuristics. This research question aims to evaluate the impact of this interleaving strategy compared to the seed selection based on a single source. 
\end{description}

To answer these questions, we follow the recommended experimental design for fuzzing experiments \cite{klees2018evaluating, bohme2022reliability}.

\textit{Benchmark.} We selected the subjects from \profuzzbench \cite{profuzzbench} as our benchmark. \profuzzbench is a widely-used benchmarking platform for evaluating stateful fuzzers of network protocols \cite{chatafl, nyx-net, nsfuzz, stateafl}. \profuzzbench comprises a suite of mature and open-source programs that implement well-known network protocols (e.g., \texttt{SSH} and \texttt{FTP}). In addition, it integrates a set of protocol fuzzing tools, including \aflnet. Our experiments were conducted on \profuzzbench using the default versions of these subjects. 

\textit{Performance Metrics and Measures.} For each experiment, we report both code coverage and state space coverage. The key idea is that a bug cannot be exposed in uncovered code or states. To evaluate \emph{code coverage}, we measure the branch coverage achieved using the automated tooling provided by the benchmarking platform \profuzzbench \cite{profuzzbench}. To evaluate the coverage of the state space, we measure the number of state transitions constructed in the protocol state machine IPSM. Additionally, we use the Vargha-Delaney effect size (${{\hat{A}_{12}}}$) to measure the statistical significance of the comparison results between two independent groups, which in our case are \aflnet and its variants.

\textit{Experimental Configuration and Infrastructure.} \changes{We conducted all the experiments using the Git commit \texttt{62d63a5} of \aflnet. The SHIFT\_SIZE parameter was set to half of the bitmap size, reserving one-half of the bitmap space for state coverage and the other half for code coverage. We set STATE\_SIZE to 256 based on our observations from our preliminary experiments.} All experiments were run on an Intel{\textregistered} Xeon{\textregistered} Platinum 8468V CPU with 192 logical cores clocked at 2.70GHz, 512GB of memory, and running Ubuntu 22.04.3 LTS. Each experiment runs for 24 hours.  We report the average results over 10 runs to mitigate the impact of randomness.

\subsection*{RQ.1 Effectiveness of State Feedback alone}

To evaluate the effectiveness of state feedback alone in guiding the fuzzing campaign, we developed two additional invariant tools of \aflnet for comparison: 

\begin{itemize}
\item \aflnetDark: a dark-box version of \aflnet with only state feedback enabled,
\item \aflnetBlack: a black-box version of \aflnet with both code and state feedback disabled.
\end{itemize}
In this experiment, \aflnetDark is our focus, \aflnetBlack serves as the baseline, and \aflnet represents the target that \aflnetDark aims to achieve.

\begin{table*}[t]
    \setlength{\abovecaptionskip}{5pt}%
    \caption{Average branch coverage and average state coverage across 10 runs of 24 hours achieved by \aflnetQueue compared to \aflnetCode. }
    \label{tab:coverage_rq2}
    \centering
    \small
    \setlength\tabcolsep{4pt}
    \def\arraystretch{1.1}
     \begin{tabular}{l|rrrr|rrrr}
        \toprule
        \multirow{2}{*}{\bfseries Subject} & \multicolumn{4}{c|}{\bfseries Code Coverage Comparison} &  \multicolumn{4}{c}{ \bfseries State Coverage Comparison}  \\ 
        \cline{2-9} 
         &  { \changes{\aflnetCodeBold}} & { \changes{\aflnetQueueBold}} & {\bfseries Improv} &  {\bfseries \footnotesize ${\boldsymbol{\hat{A}_{12}}}$ } &  { \changes{\aflnetCodeBold}} &{\changes{\aflnetQueueBold}} & { \bfseries Improv} &   {\bfseries \footnotesize ${\boldsymbol{\hat{A}_{12}}}$ } \\
        \hline
        \hline
        Bftpd  & 484.8 & 491.5 & +1.38\%	& 0.72 	& 170.5 & 334.0	& +0.96$\times$ &	1.00\\
        DCMTK  & 3076.6 & 3086.3 & +0.32\%	& 0.56 & 3.0	& 3.0	& 0.00$\times$	& 0.50 \\
        DNSmasq  & 1221.0 & 1217.6 & -0.28\%	& 0.43 	& 282.5	& 27364.0 & +95.85$\times$ &	1.00 \\
        Exim  & 2847.7 & 2862.7 & +0.53\%	& 0.46 	& 61.6	&	75.7 & +0.23$\times$ &	1.00	\\
        forked-daapd  & 2384.9 & 2401.3 & +0.69\%	& 0.54 	& 22.0 & 37.7	& +0.71$\times$ &	1.00\\
        Kamailio  & 9800.2 & 9752.4 & -0.49\%	& 0.43 	& 89.2 & 300.3	& +2.37$\times$	& 1.00 \\
        LightFTP  & 355.8 & 347.2 & -2.42\%	& 0.21 	& 179.0	& 388.7 & +1.17$\times$ &	1.00\\
        Live555  & 2809.8 & 2818.5 & +0.31\%	& 0.53 	& 75.1	& 87.9 & +0.17$\times$	& 1.00 \\
        OpenSSH  & 3336.7 & 3300.0 & -1.10\%	& 0.20 & 93.5	& 30480.9  & +325.00$\times$	& 1.00 \\
        ProFTPD  & 5296.4 & 5309.6 & +0.25\%	& 0.55 	& 250.6	& 473.5 & +0.89$\times$	& 1.00 \\
        Pure-FTPd  & 1268.0 & 1277.1 & +0.72\%	& 0.45  &	292.1	& 420.2 & +0.44$\times$ &	1.00\\
        TinyDTLS & 574.4  & 575.7 & +0.23\%	& 0.50  &	30.5 & 37.5	& +0.23$\times$ &	1.00 \\
        \hline
        \hline
        \bfseries Average & - & - & +0.01\% & - & - & - & +35.67$\times$ & - \\
        \bottomrule
    \end{tabular}
\end{table*} 

\autoref{fig:coverage-trend} shows the trends in average code coverage over time for \aflnet, \aflnetDark and \aflnetBlack. Overall, state feedback alone had no negative impact on code coverage across all subjects. With the guidance from state feedback, \aflnetDark significantly outperformed \aflnetBlack in terms of code coverage in 6 of the 12 \profuzzbench subjects (i.e., Bftpd, DNSmasq, Kamailio, LightFTP, ProFTPD and Pure-FTPd). In particular, in the subject Bftpd, \aflnetDark even performed similarly to \aflnet. In the subjects OpenSSH and TinyDTLS, although \aflnetDark only slightly improved the number of code branches covered at the end of the fuzzing campaign, it achieved the same branch number approximately 6$\times$ and 4$\times$ faster than \aflnetBlack, respectively, which can obviously reduce the fuzzing time. Unfortunately, for other subjects (i.e., DCMTK, Exim, forked-daapd, and Live555), there was almost no difference in code coverage between \aflnetDark and \aflnetBlack. 

To investigate the reason for this difference, we collected the state number of all subjects at the end of the fuzzing campaign. It is interesting to note that for the subjects where \aflnetDark did not outperform \aflnetBlack, the state numbers were also lower compared to those where \aflnetDark showed better performance. For example, the state number of the subject DCMTK is 3, while there are 334 states observed for the subject Bftpd. This result is expected, as an insufficient number of states is inadequate for guiding the fuzzing campaign, leading to a similar performance between \aflnetDark and \aflnetBlack. Therefore, we conclude that state feedback alone is effective in guiding the fuzzing campaign, provided there is a reasonable number of states to offer guidance. When code feedback is unavailable, state feedback is a fallback guidance for improving code coverage. 

\result{State feedback alone is effective in guiding the fuzzing campaign \textit{when} the state number is reasonable. }

\subsection*{RQ.2 Effectiveness of Additional State Feedback}

In this experiment, we examined whether a fuzzer with additional state guidance could outperform one with only code guidance. For this purpose, we compared two variants of \aflnet: 
\begin{itemize}
\item \aflnetCode: a variant \aflnet with only code feedback,
\item \aflnetQueue: a variant \aflnet with both code and state feedback.
\end{itemize}
Both \aflnetQueue and \aflnetCode select interesting seeds based on the test order in the corpus queue. We did not include the original \aflnet for \changes{comparison because} it uses an interleaving strategy between the seed-queue order and the state heuristics to select interesting seeds. Since this experiment specifically focuses on the impact of additional state guidance, we developed the invariant fuzzer \aflnetQueue to eliminate the influence of seed-selection strategies.

\autoref{tab:coverage_rq2} shows the average number of code branches and states covered by \aflnetQueue and \aflnetCode across all subjects. To quantify the improvement of \aflnetQueue over \aflnetCode, we report the percentage improvements in terms of code coverage and state coverage achieved in 24 hours, respectively (\textit{Improv}), as well as the possibility that a random campaign of \aflnetQueue outperforms a random campaign of \aflnetCode (${{\hat{A}_{12}}}$). We consider the Vargha-Delaney effect size ${{\hat{A}_{12}}} \geq 0.71$ or ${{\hat{A}_{12}}} \leq 0.29$ to indicate a substantial advantage of \aflnetQueue over \aflnetCode, or vice versa.  

In the aspect of code coverage, the additional state feedback in \aflnetQueue had a mixed impact when compared to \aflnetCode, which uses only code feedback. \aflnetQueue outperformed \aflnetCode in 8 out of 12 subjects, with only 1 subject (i.e., LightFTP) showing statistically significant improvement. In contrast, for the remaining 4 subjects, the additional state feedback had a negative impact on the code coverage, although only in the subjects LightFTP and OpenSSH in a statistically significant way. Overall, while additional state coverage can slightly improve code coverage for most subjects, this improvement is not statistically significant (7 out of the 12 subjects). 

In the aspect of state coverage, \aflnetQueue covered 35.67$\times$ more states on average. For nearly all subjects (except the subject DCMTK), the Vargha-Delaney effect size ${{\hat{A}_{12}}} = 1.00$ indicates a substantial advantage of \aflnetQueue over \aflnetCode in exploring state space. In addition, it is worth noting that some subjects (e.g., DCMTK and forked-daapd) only exhibited a small number of observed states. It is expected that this sparse feedback is not effective in improving the code coverage. Conversely, the results for DNSmasq and OpenSSH suggest that overly dense feedback can also be ineffective. Overall, there is no correlation between the number of states observed and the improvements in code coverage.

\begin{table*}[t]
    \setlength{\abovecaptionskip}{5pt}%
    \caption{Average branch coverage across 10 runs of 24 hours achieved by \aflnet compared to \aflnetQueue and \aflnetIPSM. }
    \label{tab:code_coverage_rq3}
    \centering
    \small
    \def\arraystretch{1.1}
     \begin{tabular}{l|r|rrr|rrr}
        \toprule
        \multirow{2}{*}{\bfseries Subject} & \multirow{2}{*}{\bfseries \aflnetBold} & \multicolumn{3}{c|}{\bfseries Comparison with \aflnetQueueBold} &  \multicolumn{3}{c}{ \bfseries Comparison with \aflnetIPSMBold}  \\ 
        \cline{3-8} 
         &  & { \aflnetQueueBold} & {\bfseries Improv} &  {\bfseries \footnotesize ${\boldsymbol{\hat{A}_{12}}}$ } 
         & { \aflnetIPSMBold} & { \bfseries Improv} &   {\bfseries \footnotesize ${\boldsymbol{\hat{A}_{12}}}$ } \\
        \hline
        \hline
        Bftpd & 487.0 & 491.5	& -0.92\%	& 0.23 & 486.6 &	+0.08\%	& 0.47 \\
        DCMTK & 3120.4 & 3086.3	& +1.10\%	& 0.95 & 3113.7 &	+0.22\%	& 0.70 \\
        DNSmasq & 1202.5 & 1217.6	&-1.24\%	& 0.00	&1194.9	& +0.64\%	& 0.70 \\
        Exim & 2922.0 	& 2862.7	& +2.07\%	& 0.36	& 2888.8	& +1.15\%	& 0.80 \\
        forked-daapd & 2329.9 & 2401.3	&-2.97\%	& 0.13	& 2279.4	& +2.22\%	& 0.80  \\
        Kamailio & 9899.4  & 9752.4 &	+1.51\%	& 0.97	& 9824.6	& +0.76\%	& 0.68 \\
        LightFTP & 346.6 & 347.2	&-0.16\%	& 0.33	& 345.5 &	+0.32\%	& 0.62 \\
        Live555 & 2808.1 & 2818.5	&-0.37\%	& 0.41	& 2780.5&	+0.99\%	& 0.75 \\
        OpenSSH & 3353.8 & 3300.0	& +1.63\%	& 0.82	& 3341.4 &	+0.37\%	& 0.65 \\
        ProFTPD & 5324.2 & 5309.6	& +0.27\%	& 0.58	& 5150.5	& +3.37\%	& 0.89\\
        Pure-FTPd & 1167.5 & 1277.1	& -8.58\%	& 0.00	& 1075.4	& +8.56\%	& 0.96 \\
        TinyDTLS & 583.8 & 575.7	& +1.41\%	& 0.79	& 577.5	& +1.09\%	& 0.76 \\
        \hline
        \hline
        \bfseries Average & - & - & -0.52\% & - & - & +1.65\% & - \\
        \bottomrule
    \end{tabular}
\end{table*} 

\begin{table*}[t]
    \setlength{\abovecaptionskip}{5pt}%
    \caption{Average state coverage across 10 runs of 24 hours achieved by \aflnet compared to \aflnetQueue and \aflnetIPSM. }
    \label{tab:state_coverage_rq3}
    \centering
    \small
    \def\arraystretch{1.1}
     \begin{tabular}{l|r|rrr|rrr}
        \toprule
        \multirow{2}{*}{\bfseries Subject} & \multirow{2}{*}{\bfseries \aflnetBold} & \multicolumn{3}{c|}{\bfseries Comparison with \aflnetQueueBold} &  \multicolumn{3}{c}{ \bfseries Comparison with \aflnetIPSMBold}  \\ 
        \cline{3-8} 
         &  & { \aflnetQueueBold} & {\bfseries Improv} &  {\bfseries \footnotesize ${\boldsymbol{\hat{A}_{12}}}$ } 
         & { \aflnetIPSMBold} & { \bfseries Improv} &   {\bfseries \footnotesize ${\boldsymbol{\hat{A}_{12}}}$ } \\
        \hline
        \hline
        Bftpd & 334.3 & 334.0	& +0.09\%	& 0.57	& 335.0 &	-0.21\%	& 0.37 \\
        DCMTK & 3.0 & 3.0	& 0.00\%	& 0.50	& 3.0	& 0.00\%	& 0.50 \\
        DNSmasq & 32256.5 & 27364.0 &	+17.88\%	& 1.00	& 26982.2	& +19.55\%	& 1.00\\
        Exim & 69.1 & 75.7 &	-8.72\%	& 0.45	& 66.0	& +4.70\%	& 0.51\\
        forked-daapd &  43.2 & 37.7	& +14.67\%	& 1.00	& 39.0	& +10.85\%	& 1.00\\
        Kamailio & 313.0 & 300.3	& +4.23\%	& 0.89	& 235.2	& +33.10\%	& 0.70\\
        LightFTP & 380.4 & 388.7	& -2.14\%	& 0.46	& 375.9	& +1.20\%	& 0.58\\
        Live555 & 91.7 & 87.9	& +4.32\%	& 0.57	& 89.3	& +2.69\%	& 0.69\\
        OpenSSH & 35433.5 & 30480.9	& +16.25\%	& 1.00	& 30943.6	& +14.51\%	&  1.00 \\
        ProFTPD & 476.0 & 473.5	& +0.53\%	& 0.59 &	359.9 &	+32.26\% &	1.00\\
        Pure-FTPd & 521.2 & 420.2 &	+24.05\%	& 1.00	& 463.8	& +12.39\%	& 1.00\\
        TinyDTLS & 36.8 & 37.5 &	-1.87\%	& 0.46	& 30.1	& +22.26\%	& 0.74 \\
        \hline
        \hline
        \bfseries Average & - & - & +5.77\% & - & - & +12.77\% & - \\
        \bottomrule
    \end{tabular}
\end{table*} 

These experimental results demonstrate that additional state feedback can effectively guide the fuzzer to explore more states. \changes{The} additional state guidance did show significant effectiveness in improving the code coverage, \changes{and} it also has no obvious harm on most subjects (i.e., 10 out of 12 subjects). A possible explanation is that \aflnet considers response codes from a server as the representation of states; however, this may not be a good state definition for each subject, as noted by the follow-up works of \aflnet \cite{sgfuzz, nsfuzz}.

\result{Additional state feedback, as defined by \aflnet, can effectively guide the fuzzer to explore a larger state space. However, it does not result in significant improvement in code coverage.}

\subsection*{RQ.3 Impact of Seed-Selection Strategies}

To evaluate the impact of seed-selection strategies, we compare \aflnet with two alternative implementations:
\begin{itemize}
\item \aflnetQueue: a variant \aflnet that selects interesting seeds \textit{only} based on the order of the seed queue,
\item \aflnetIPSM: a variant \aflnet that selects interesting seeds \textit{only} based on the state heuristics. 
\end{itemize}
\aflnet selects interesting seeds using an interleaving strategy between the order of the seed queue and the state heuristics. All these tools are configured with both code- and state- feedback. We compare three tools in terms of the code coverage and the state coverage. 

\autoref{tab:code_coverage_rq3} shows the average code branches covered by \aflnet, \aflnetQueue and \aflnetIPSM across 10 runs of 24 hours. In addition, we report the improvement in the code coverage of \aflnet compared to \aflnetQueue and \aflnetIPSM, respectively (\emph{Improv}), and the Vargha-Delaney effect size (${\hat{A}_{12}}$). Compared to the seed-selection strategy based on queue order (i.e.,\aflnetQueue), the interleaving seed-selection strategy (i.e., \aflnet) performed significantly better in some subjects (i.e., DCMTK, Kamailio, OpenSSH, and TinyDTLS). However, it underperformed in the subjects DNSmasq, forked-daapd, and Pure-FTPd. In the remaining subjects, both seed-selection strategies had similar performance. Compared to the strategy based on the state heuristics (i.e., \aflnetIPSM), the interleaving strategy consistently performed better although the improvement was not statistically significant in some subjects (e.g., Bftpd). 

\autoref{tab:state_coverage_rq3} shows the state coverage of \aflnet, \aflnetQueue and \aflnetIPSM in a similar format to our previous table. Overall, \aflnet outperformed both baseline tools \aflnetQueue and \aflnetIPSM in terms of state coverage. \aflnet covered 5.77\% more states than \aflnetQueue, and 12.77\% more states than \aflnetIPSM on average. Although \aflnet covers fewer states than \aflnetIPSM or \aflnetQueue in the subjects Bftpd, LightFTP, and TinyDTLS, this underperformance is not significant given the values of ${{\hat{A}_{12}}}$. 

Considering both aspects of code coverage and state coverage, \aflnet is the best performer among the comparison tools across all subjects. Therefore, the interleaving seed-selection strategy is generally the best configuration while testing most subjects. However, if the primary goal is to maximize code coverage, users might consider configuring the fuzzer with a seed-selection strategy based solely on queue order, as suggested by the comparison between \aflnet and \aflnetQueue. On the other hand, if maximizing state coverage is the objective, the interleaving seed-selection strategy is undoubtedly the best choice. 

In addition, it is interesting to note that there is no obvious correlation between state coverage and code coverage. A fuzzer that covers more states does not necessarily mean it would cover more code as well, as demonstrated in some subjects (e.g., DNSmasq, forked-daapd, and Pure-FTPd). 

\result{The fuzzer configured with interleaving seed-selection strategy outperforms selecting interesting seeds based only on queue order or state heuristics. }
\section{Recent Progress in Stateful Fuzzing}

\aflnet has significantly advanced fuzzing techniques for network protocols. During the fuzzing campaign, \aflnet acts as the client application, establishing real network connections with the server under test and then exchanging messages. This approach follows the real-world architectures of network protocols, reducing the manual effort of understanding network protocols and modifying source codes. \aflnet is widely regarded as an optimal choice for network protocol fuzzing \footnote{\url{https://github.com/AFLplusplus/AFLplusplus/blob/stable/docs/best_practices.md\#fuzzing-a-network-service}}, and has uncovered numerous critical vulnerabilities in widely-used protocol implementations. However, it also has several shortcomings that the research community has actively addressed, leading to enhancements in various aspects. 

\inlineSubSubSection{What is a state?}
Considering state feedback and optimizing state coverage is a key contribution of \aflnet. Yet, what do we consider as the current ``state'' and how do we identify it? In the default setting, \aflnet uses the response code extracted from the response message to represent the current protocol state. However, the response code is a very coarse representation of states. As demonstrated in our previous experiment, this state representation does not significantly improve code coverage. In addition, the response codes are not always available in response messages. 

To address this limitation, a series of works have proposed alternative state representations. 
\sgfuzz \cite{sgfuzz} uses the sequence of values assigned to state variables (identified as \textit{enum} type variables) to represent the sequence of protocol states corresponding to a message sequence.
Similarly, \nsfuzz \cite{nsfuzz} introduces a variable-based state representation to infer states of network protocols.
\stateafl \cite{stateafl} infers protocol states by taking snapshots of long-lived memory areas.
Utilizing the recent advances in large language models (LLMs), \chatafl \cite{chatafl} uses LLMs to infer states based on exchanged messages.
These approaches effectively address the challenge of state identification in \aflnet.

\inlineSubSubSection{How to maximize state coverage?}
\aflnet provides three seed-selection (line 10 of \autoref{alg:sgf}) algorithms: \texttt{FAVOR}, \texttt{RANDOM}, and \texttt{ROUND-ROBIN}, with \texttt{FAVOR} being the default configuration. The \texttt{FAVOR} algorithm prioritizes states that are rarely exercised, giving them more chances to be tested. The \texttt{RANDOM} algorithm selects states randomly, while the \texttt{ROUND-ROBIN} algorithm maintains states in a circular queue and selects them in turns. However, as shown by Liu et al. \cite{aflnetlegion}, these three algorithms yield similar results in terms of code coverage. 

Subsequent works have sought to propose more principled approaches to state selection.
Borcherding et al. \cite{banditstate} model state selection as a Multi-armed Bandit Problem. Unfortunately, the authors found that this approach prevents the fuzzer from reaching deeper states, resulting in worse code coverage compared to \aflnet. \aflnetLegion \cite{aflnetlegion} introduces a novel seed-selection algorithm to \aflnet based on \legion \cite{legion}, a variant of Monte Carlo tree search. However, the performance improvements of \aflnetLegion turn out to be not statistically significant. We believe that this is explained by the low fuzzing throughput of baseline \aflnet, which hinders the full potential of this systematic algorithm.
Once this challenge is resolved, it is worthwhile to explore other heuristics that have shown promise in (code) coverage-guided greybox fuzzing \cite{aflfast,fairfuzz,aflgo}.

\inlineSubSubSection{How to maximize fuzzing throughput?}
\aflnet operates with low fuzzing throughput, averaging around 20 executions per second, due to several factors: Firstly, \aflnet sends inputs through the network sockets, which is significantly slower than reading inputs from files. Secondly, it introduces a time delay between messages to ensure the server is ready to receive the next message. Lastly, it runs a clean-up script to reset the state of the environment after each iteration. 

\greenfuzz \cite{greenfuzz} improves the fuzzing throughput of \aflnet by utilizing a simulated socket library \texttt{Desock+}, a modified version of \texttt{preeny}, to reduce system call overhead.
\nyxnet \cite{nyx-net} employs hypervisor-based snapshot fuzzing to ensure noise-free execution and accelerate state resets.
\snapfuzz \cite{snapfuzz} enhances the fuzzing throughput by introducing several strategies, including transforming slow asynchronous network communication into fast synchronous communication, snapshotting states, and using in-memory filesystems.
These approaches significantly increase the fuzzing throughput of \aflnet.

\inlineSubSubSection{How to maximize the syntactic validity of each message?} 
During message mutation, \aflnet uses the same byte-level mutation operators as traditional greybox fuzzers, which can easily break the structure of a valid message. In principle, existing grammar-aware strategies can be applied to \aflnet to improve the effectiveness of message mutation. Given a user-provided data model describing the message grammar, structure-aware blackbox protocol fuzzers \cite{peach, bestorm} generate valid messages from scratch, while structure-aware greybox fuzzers \cite{aflsmart,nautilus} take a mutation-based approach.
In contrast, \chatafl \cite{chatafl} obtains the message structure information from the LLMs and then preserves valid message grammar during mutation. In addition, there are several existing works \cite{polyglot, grimoire, superion} that dynamically infer message structures based on the observed messages. We can distinguish blackbox approaches \cite{learningandFuzz,protolearn} that learn the message structure from a given corpus of messages and whitebox approaches \cite{polyglot,tupni} that actively explore the protocol implementation to uncover message structure. For instance, Polyglot \cite{polyglot} uses dynamic analysis techniques, such as tainting and symbolic execution, to extract the message format from the protocol implementation.

\inlineSubSubSection{Protocol Environment Fuzzing.}
\aflnet focuses on only fuzzing network traffic over a specific port. However, beyond this single input source, network protocols often interact with complex execution environments such as configuration files, databases, and other network sockets, which can affect the behaviors as well.
\chaosafl \cite{chaosafl} involves all file-related inputs as fuzzing targets, while \envfuzz \cite{envfuzz} considers the full program environment in the system-call layer as fuzzing targets. Both approaches extend the capability of \aflnet in finding environment-inducing bugs.

\section{Reflections and Path Forward}

Over the past five years, \aflnet{} has made significant contributions to research, practice, and education. In terms of research impact, the short tool demo paper of \aflnet{} has been cited over 270 times (as of November 2024, according to Google Scholar), with many citations appearing in premier conferences and journals in Security and Software Engineering. Regarding practical impact, \aflnet{} has garnered 872 stars on GitHub and currently supports 17 protocols, 12 of which were contributed by other researchers, demonstrating its versatility and community engagement. 

Security researchers have also published experience reports and tutorials on using \aflnet{} for challenging targets \cite{nccgroup5g,dicomaflnet,etaswebinar,matteraflnetbug1,matteraflnetbug2}. For example, the NCC Group explored the challenges of fuzzing 5G protocols \cite{nccgroup5g} and demonstrated \aflnet{}'s ability to uncover bugs in this critical domain\footnote{NCC Group reported that \aflnet{} identified some crash-triggering issues, which were under further investigation and subject to coordinated disclosure as appropriate.}. Similarly, researchers from the University of Melbourne extended \aflnet{} to support IPv6 for fuzz testing the software development kit (SDK) of Matter, a novel application-layer protocol designed to unify fragmented smart home ecosystems \cite{matterprotocol}. This extension has discovered zero-day vulnerabilities in the Matter SDK \cite{matteraflnetbug1, matteraflnetbug2}. Moreover, ETAS, a subsidiary of Robert Bosch GmbH, highlighted \aflnet{} as a potential open-source protocol fuzzing solution in the context of the ISO/SAE 21434 standard for road vehicle cybersecurity engineering \cite{etaswebinar}. In education, \aflnet{} has been introduced to hundreds of Master's students through modules such as "Security and Software Testing (SWEN90006)" \cite{swen90006} at the University of Melbourne and "Fantastic Bugs and How to Find Them (17-712)" \cite{cmucourse} at Carnegie Mellon University.

Why has our work on AFLNet generated such practical and academic impact in a short period of \changes{fewer than} five years? We can see two reasons: (i)~our open science approach and (ii)~providing a practical solution to a long-standing problem of validating reactive systems. As for our \emph{open science} approach, we strongly believe that sound progress in science requires reproducibility and that effective impact in practice requires open source. AFLNet is an excellent case demonstrating the success of our open science approach. Today, it is expected that the tools and artifacts are published together with the paper. Five years ago, it was not common to make prototypes publicly available as open source \cite{sok}.

AFLNet is a practical solution to the long-standing problem of validating reactive systems. Looking back and reflecting on it, we feel this is because of the sheer dearth of suitable approaches for testing reactive systems, though there exist many approaches for testing sequential transformational systems. Reactive systems are in continuous interaction with the environment by exchanging messages or events between the system and the environment. Thus the ``input" to a reactive system is not a single event but rather a sequence of events. Most protocol implementations are reactive systems - instead, the sequence of messages that can be legitimately exchanged is the so-called protocol!  Prior to the greybox approach of AFLNet, reactive system validation would typically need to be carried out via stateful blackbox fuzzing approaches or via algorithmic whitebox verification approaches such as model checking.  We already mentioned the deficiencies of using stateful blackbox fuzzing approaches - since they involve manual writing of a state model and data model. Moreover, the effectiveness of the stateful blackbox fuzzing approach depends on how complete the manually\changes{-}written state model and data model are.

Regarding the use of model checking for validating reactive systems, this would suffer from various limitations. 
\begin{itemize}
\item \changes{A} temporal logic property needs to be provided to guide the validation exercise via model checking.
\item \changes{The} validation will be carried out at the model level where only the protocol model is being checked. Alternatively, if the protocol implementation is being checked, an abstraction will need to be designed to extract a finite state transition system from the infinite state protocol implementation, as per the abstraction-refinement approach of software model checking \cite{slam}. 
\item \changes{Finally,} after a bug is found, it is reported in the form of a counter-example trace from where a buggy event sequence still needs to be extracted. 
\end{itemize}

The work of AFLNet and subsequent works free the practitioner from all of these steps, thereby constituting a significant practical advance. It also represents a significant conceptual and practical advance over greybox fuzzing by accompanying greybox fuzzing with lightweight model learning. This has opened up the applicability of greybox fuzzing from stateless systems like file format parsers to a plethora of stateful, reactive applications. Recent works in the research community on extending greybox fuzzing to concurrent and distributed systems (e.g., \cite{ccs23, asplos24}) also rely partially on the core advance in the AFLNet work. Moving forward, we may thus see a much wider variety of stateful, reactive, concurrent, distributed application software being routinely checked via greybox fuzzing. These advances would constitute the broader and longer-term impact of the AFLNet work in 2020.

\section*{Acknowledgments}
This research is partially supported by the National Research Foundation, Singapore, and Cyber Security Agency of Singapore under its National Cybersecurity R\&D Programme (Fuzz Testing NRF-NCR25-Fuzz-0001). Any opinions, findings and conclusions, or recommendations expressed in this material are those of the author(s) and do not reflect the views of National Research Foundation, Singapore, and Cyber Security Agency of Singapore. This research is also partially funded by the European Union. Views and opinions expressed are however those of the author(s) only and do not necessarily reflect those of the European Union or the European Research Council Executive Agency. Neither the European Union nor the granting authority can be held responsible for them. This work is supported by ERC grant (Project AT SCALE, 101179366).

\bibliographystyle{IEEEtran}
\bibliography{references}

\begin{thebibliography}{10}
\providecommand{\url}[1]{#1}
\csname url@samestyle\endcsname
\providecommand{\newblock}{\relax}
\providecommand{\bibinfo}[2]{#2}
\providecommand{\BIBentrySTDinterwordspacing}{\spaceskip=0pt\relax}
\providecommand{\BIBentryALTinterwordstretchfactor}{4}
\providecommand{\BIBentryALTinterwordspacing}{\spaceskip=\fontdimen2\font plus
\BIBentryALTinterwordstretchfactor\fontdimen3\font minus \fontdimen4\font\relax}
\providecommand{\BIBforeignlanguage}[2]{{%
\expandafter\ifx\csname l@#1\endcsname\relax
\typeout{** WARNING: IEEEtran.bst: No hyphenation pattern has been}%
\typeout{** loaded for the language `#1'. Using the pattern for}%
\typeout{** the default language instead.}%
\else
\language=\csname l@#1\endcsname
\fi
#2}}
\providecommand{\BIBdecl}{\relax}
\BIBdecl

\bibitem{peach}
\BIBentryALTinterwordspacing
``Peach {Fuzzer} {Platform}.'' [Online]. Available: \url{https://gitlab.com/gitlab-org/security-products/protocol-fuzzer-ce}
\BIBentrySTDinterwordspacing

\bibitem{spike}
\BIBentryALTinterwordspacing
``{SPIKE Fuzzer Platform}.'' [Online]. Available: \url{{http://www.immunitysec.com}}
\BIBentrySTDinterwordspacing

\bibitem{aflsmart}
V.-T. Pham, M.~B{\"o}hme, A.~E. Santosa, A.~R. C{\u{a}}ciulescu, and A.~Roychoudhury, ``Smart greybox fuzzing,'' \emph{IEEE Transactions on Software Engineering}, vol.~47, no.~9, pp. 1980--1997, 2019.

\bibitem{aflgo}
M.~B\"{o}hme, V.-T. Pham, M.-D. Nguyen, and A.~Roychoudhury, ``Directed greybox fuzzing,'' in \emph{Proceedings of the 2017 ACM SIGSAC Conference on Computer and Communications Security}, 2017, pp. 2329--2344.

\bibitem{aflgroup}
Website, ``{AFL} user group,'' \url{https://groups.google.com/forum/#!forum/afl-users}, 2019, accessed: 2019-08-15.

\bibitem{aflnet}
V.-T. Pham, M.~B{\"o}hme, and A.~Roychoudhury, ``{AFLNet}: a greybox fuzzer for network protocols,'' in \emph{Proceedings of the 13th IEEE International Conference on Software Testing, Validation and Verification}, 2020, pp. 460--465.

\bibitem{nccgroup5g}
N.~Group, ``The challenges of fuzzing {5G} protocols,'' \url{https://www.nccgroup.com/au/research-blog/the-challenges-of-fuzzing-5g-protocols/}, 2021, accessed: 2024-10-03.

\bibitem{matteraflnetbug1}
C.~S. Alliance, ``A use-after-free vulnerability discovered by {AFLNet},'' \url{https://github.com/project-chip/connectedhomeip/pull/33148/}, 2024, accessed: 2024-10-03.

\bibitem{matteraflnetbug2}
------, ``A critical memory (heap) leak vulnerability discovered by {AFLNet},'' \url{https://github.com/project-chip/connectedhomeip/pull/32970/}, 2024, accessed: 2024-10-03.

\bibitem{dicomaflnet}
M.~Nedyak, ``How to hack medical imaging applications via {DICOM},'' Hack In The Box Security Conference, 2020, accessed: 2024-10-03.

\bibitem{etaswebinar}
E.~GmbH, ``Demystifying a current trend - security fuzz testing in the context of {ISO/SAE} 21434,'' \url{https://youtu.be/HDfkD67UUSw}, 2024, accessed: 2024-10-03.

\bibitem{profuzzbench}
R.~Natella and V.-T. Pham, ``{ProFuzzBench}: A benchmark for stateful protocol fuzzing,'' in \emph{Proceedings of the 30th ACM SIGSOFT international symposium on software testing and analysis}, 2021, pp. 662--665.

\bibitem{afl}
\BIBentryALTinterwordspacing
M.~Zalewski, ``{AFL}.'' [Online]. Available: \url{https://lcamtuf.coredump.cx/afl/}
\BIBentrySTDinterwordspacing

\bibitem{aflfast}
M.~B{\"o}hme, V.-T. Pham, and A.~Roychoudhury, ``Coverage-based greybox fuzzing as markov chain,'' in \emph{Proceedings of the 2016 ACM SIGSAC Conference on Computer and Communications Security}, 2016, pp. 1032--1043.

\bibitem{libfuzzer}
\BIBentryALTinterwordspacing
``{libFuzzer} – a library for coverage-guided fuzz testing,'' {LLVM}. [Online]. Available: \url{https://llvm.org/docs/LibFuzzer.html}
\BIBentrySTDinterwordspacing

\bibitem{bestorm}
\BIBentryALTinterwordspacing
``{beSTORM} black box testing.'' [Online]. Available: \url{https://www.beyondsecurity.com/bestorm.html}
\BIBentrySTDinterwordspacing

\bibitem{boofuzz}
\BIBentryALTinterwordspacing
Jtpereyda, ``Boofuzz: A fork and successor of the sulley fuzzing framework.'' [Online]. Available: \url{https://github.com/jtpereyda/boofuzz}
\BIBentrySTDinterwordspacing

\bibitem{sulley}
\BIBentryALTinterwordspacing
``Sulley: A pure-python fully automated and unattended fuzzing framework.'' [Online]. Available: \url{{https://github.com/OpenRCE/sulley}}
\BIBentrySTDinterwordspacing

\bibitem{lightftp}
\BIBentryALTinterwordspacing
``{LightFTP} {Server}.'' [Online]. Available: \url{https://github.com/hfiref0x/LightFTP}
\BIBentrySTDinterwordspacing

\bibitem{ftp}
IEFT, ``{File} {Transfer} {Protocol}.'' \url{https://tools.ietf.org/html/rfc959}, 1985, accessed: 2019-08-12.

\bibitem{aflbugs}
\BIBentryALTinterwordspacing
``{AFL} vulnerability trophy case.'' [Online]. Available: \url{http://lcamtuf.coredump.cx/afl/\#bugs}
\BIBentrySTDinterwordspacing

\bibitem{klees2018evaluating}
G.~Klees, A.~Ruef, B.~Cooper, S.~Wei, and M.~Hicks, ``Evaluating fuzz testing,'' in \emph{Proceedings of the 2018 ACM SIGSAC conference on computer and communications security}, 2018, pp. 2123--2138.

\bibitem{bohme2022reliability}
M.~B{\"o}hme, L.~Szekeres, and J.~Metzman, ``On the reliability of coverage-based fuzzer benchmarking,'' in \emph{Proceedings of the 44th International Conference on Software Engineering}, 2022, pp. 1621--1633.

\bibitem{chatafl}
R.~Meng, M.~Mirchev, M.~B{\"o}hme, and A.~Roychoudhury, ``Large language model guided protocol fuzzing,'' in \emph{Proceedings of the 31st Annual Network and Distributed System Security Symposium}, 2024.

\bibitem{nyx-net}
S.~Schumilo, C.~Aschermann, A.~Jemmett, A.~Abbasi, and T.~Holz, ``Nyx-net: network fuzzing with incremental snapshots,'' in \emph{Proceedings of the 17th European Conference on Computer Systems}, 2022, pp. 166--180.

\bibitem{nsfuzz}
S.~Qin, F.~Hu, Z.~Ma, B.~Zhao, T.~Yin, and C.~Zhang, ``{NSFuzz}: Towards efficient and state-aware network service fuzzing,'' \emph{ACM Transactions on Software Engineering and Methodology}, vol.~32, no.~6, pp. 1--26, 2023.

\bibitem{stateafl}
R.~Natella, ``{StateAFL}: Greybox fuzzing for stateful network servers,'' \emph{Empirical Software Engineering}, vol.~27, no.~7, p. 191, 2022.

\bibitem{sgfuzz}
J.~Ba, M.~B{\"{o}}hme, Z.~Mirzamomen, and A.~Roychoudhury, ``Stateful greybox fuzzing,'' in \emph{Proceedings of the 31st USENIX Security Symposium}.\hskip 1em plus 0.5em minus 0.4em\relax {USENIX} Association, 2022, pp. 3255--3272.

\bibitem{aflnetlegion}
D.~Liu, V.-T. Pham, G.~Ernst, T.~Murray, and B.~I. Rubinstein, ``State selection algorithms and their impact on the performance of stateful network protocol fuzzing,'' in \emph{Proceedings of the 2022 IEEE International Conference on Software Analysis, Evolution and Reengineering}.\hskip 1em plus 0.5em minus 0.4em\relax IEEE, 2022, pp. 720--730.

\bibitem{banditstate}
A.~Borcherding, M.~Giraud, I.~Fitzgerald, and J.~Beyerer, ``The bandit’s states: Modeling state selection for stateful network fuzzing as multi-armed bandit problem,'' in \emph{Proceedings of the 2023 IEEE European Symposium on Security and Privacy Workshops}, 2023, pp. 345--350.

\bibitem{legion}
D.~Liu, G.~Ernst, T.~Murray, and B.~I. Rubinstein, ``Legion: Best-first concolic testing,'' in \emph{Proceedings of the 35th IEEE/ACM International Conference on Automated Software Engineering}, 2020, pp. 54--65.

\bibitem{fairfuzz}
C.~Lemieux and K.~Sen, ``{FairFuzz}: A targeted mutation strategy for increasing greybox fuzz testing coverage,'' in \emph{Proceedings of the 33rd ACM/IEEE International Conference on Automated Software Engineering}, 2018, pp. 475--485.

\bibitem{greenfuzz}
S.~B. Andarzian, C.~Daniele, and E.~Poll, ``{Green-Fuzz}: Efficient fuzzing for network protocol implementations,'' in \emph{International Symposium on Foundations and Practice of Security}, 2023, pp. 253--268.

\bibitem{snapfuzz}
A.~Andronidis and C.~Cadar, ``Snapfuzz: high-throughput fuzzing of network applications,'' in \emph{Proceedings of the 31st ACM SIGSOFT International Symposium on Software Testing and Analysis}, 2022, pp. 340--351.

\bibitem{nautilus}
C.~Aschermann, T.~Frassetto, T.~Holz, P.~Jauernig, A.-R. Sadeghi, and D.~Teuchert, ``Nautilus: Fishing for deep bugs with grammars,'' in \emph{Proceedings of the 2019 Network and Distributed System Security Symposium}, 2019.

\bibitem{polyglot}
J.~Caballero, H.~Yin, Z.~Liang, and D.~Song, ``Polyglot: Automatic extraction of protocol message format using dynamic binary analysis,'' in \emph{Proceedings of the 14th ACM conference on Computer and communications security}, 2007, pp. 317--329.

\bibitem{grimoire}
T.~Blazytko, C.~Aschermann, M.~Schl{\"o}gel, A.~Abbasi, S.~Schumilo, S.~W{\"o}rner, and T.~Holz, ``{GRIMOIRE}: Synthesizing structure while fuzzing,'' in \emph{Proceedings of the 28th {USENIX} Security Symposium}, 2019, pp. 1985--2002.

\bibitem{superion}
J.~Wang, B.~Chen, L.~Wei, and Y.~Liu, ``Superion: Grammar-aware greybox fuzzing,'' in \emph{Proceedings of the IEEE/ACM 41st International Conference on Software Engineering}.\hskip 1em plus 0.5em minus 0.4em\relax IEEE, 2019, pp. 724--735.

\bibitem{learningandFuzz}
J.~Patra and M.~Pradel, ``Learning to fuzz: Application-independent fuzz testing with probabilistic, generative models of input data,'' TU Darmstadt, Tech. Rep., 2016.

\bibitem{protolearn}
R.~Fan and Y.~Chang, ``Machine learning for black-box fuzzing of network protocols,'' in \emph{Information and Communications Security}, 2018, pp. 621--632.

\bibitem{tupni}
W.~Cui, M.~Peinado, K.~Chen, H.~J. Wang, and L.~Irun-Briz, ``Tupni: Automatic reverse engineering of input formats,'' in \emph{Proceedings of the 15th ACM Conference on Computer and Communications Security}, 2008, pp. 391--402.

\bibitem{chaosafl}
Z.~Mirzamomen and M.~B{\"o}hme, ``Finding bug-inducing program environments,'' \emph{arXiv preprint arXiv:2304.10044}, 2023.

\bibitem{envfuzz}
R.~Meng, G.~J. Duck, and A.~Roychoudhury, ``Program environment fuzzing,'' in \emph{Proceedings of the 31st ACM SIGSAC Conference on Computer and Communications Security}, 2024.

\bibitem{matterprotocol}
C.~S. Alliance, ``Matter - the foundation for connected things,'' \url{https://csa-iot.org/all-solutions/matter/}, 2024, accessed: 2024-10-03.

\bibitem{swen90006}
U.~of~Melbourne, ``Security and software testing,'' \url{https://github.com/SWEN90006-2023/SWEN90006-assignment-2}, 2023, accessed: 2024-10-03.

\bibitem{cmucourse}
C.~M. University, ``Fantastic bugs and how to find them,'' \url{https://cmu-fantastic-bugs.github.io/}, 2023, accessed: 2024-10-03.

\bibitem{sok}
\BIBentryALTinterwordspacing
F.~Weissberg, J.~M\"{o}ller, T.~Ganz, E.~Imgrund, L.~Pirch, L.~Seidel, M.~Schloegel, T.~Eisenhofer, and K.~Rieck, ``Sok: Where to fuzz? assessing target selection methods in directed fuzzing,'' in \emph{Proceedings of the 19th ACM Asia Conference on Computer and Communications Security}, ser. ASIA CCS '24.\hskip 1em plus 0.5em minus 0.4em\relax New York, NY, USA: Association for Computing Machinery, 2024, p. 1539–1553. [Online]. Available: \url{https://doi.org/10.1145/3634737.3661141}
\BIBentrySTDinterwordspacing

\bibitem{slam}
T.~Ball and S.~Rajamani, ``Automatically validating temporal safety properties of interfaces,'' in \emph{International SPIN Workshop on Model Checking of Software}, 2001.

\bibitem{ccs23}
R.~Meng, G.~Pirlea, A.~Roychoudhury, and I.~Sergey, ``Greybox fuzzing of disrtibuted systems,'' in \emph{30th ACM Conference on Computer and Communications Security (CCS)}, 2023.

\bibitem{asplos24}
D.~Wolff, S.~Zheng, G.~Duck, U.~Mathur, and A.~Roychoudhury, ``Greybox fuzzing for concurrency testing,'' in \emph{29th ACM International Conference on Architectural Support for Programming Languages and Operating Systems (ASPLOS)}, 2024.

\end{thebibliography}

\end{document}